%% file: main.tex
\documentclass[conference]{IEEEtran}

\usepackage{graphicx}
\usepackage{subcaption}
\usepackage{cite}
\usepackage{amsmath,amssymb,amsfonts}
\usepackage{algorithmic}
\usepackage{textcomp}
\usepackage{xcolor}

\newcommand{\observation}[2]{%
  \par\bigskip
  \noindent
  \fbox{%
    \begin{minipage}{\dimexpr\columnwidth-2\fboxsep-2\fboxrule\relax}
      \textbf{Observation #1:} #2
    \end{minipage}
  }%
  \par\bigskip
}
\DeclareUnicodeCharacter{00AD}{}

\begin{document}

\title{Modality Inflation: Energy Characterization and Optimization Opportunities for MLLM Inference}

\author{
\IEEEauthorblockN{
Mona Moghadampanah,
Adib Rezaei Shahmirzadi,
Farhana Amin,
Dimitrios S. Nikolopoulos
}
\IEEEauthorblockA{
Department of Computer Science\\
Virginia Tech\\
Blacksburg, VA, USA\\
\{monamp, adibzi, afarhana, dsn\}@vt.edu
}
}


\maketitle


\input{sections/0-abstract}

\begin{IEEEkeywords}
Sustainable AI, Multimodal LLMs, Energy-efficient computing, GPU power management
\end{IEEEkeywords}


\input{sections/1-introduction}

\input{sections/2-background}

\input{sections/4-characterizing}

\input{sections/5-controls}

\input{sections/7-relatedWork}

\input{sections/6-futureWork}

\input{sections/9-conclusion}

\section*{Acknowledgment}
This material is based on work supported by the National Science Foundation under Grants No 2315851 and 2106634 and the Virginia Tech Foundation.

\bibliographystyle{IEEEtran}
\bibliography{refs} 

\end{document}

%% file: sections/0-abstract.tex
\begin{abstract}
    Multimodal large language models (MLLMs) are built on text-only LLMs by incorporating additional modalities, enabling multimodal understanding and a broader range of applications. However, these additions introduce a previously unexplored energy trade-off across modalities that remains poorly understood, as most prior work focuses on text-only models. In this paper, we examine modality inflation, a key source of inefficiency in which multimodal inputs increase inference workloads through extra encoding stages and expanded token sequences. We provide the first detailed, stage-level analysis of energy consumption in MLLM inference by breaking the pipeline into vision encoding, prefill, and decoding stages. Using four representative MLLMs evaluated on NVIDIA A100 GPU, we quantify the additional energy required for multimodal inference compared to text-only baselines, observing overheads ranging from 17\% to 94\% across models for identical inputs. Our results show that energy bottlenecks differ widely across model architectures, stemming either from compute-heavy vision encoders or from the downstream impact of large visual token sequences during prefill. By examining GPU power traces, we further uncover substantial GPU underutilization during multimodal execution and show that input complexity leads to markedly different energy scaling behaviors across models. Finally, we demonstrate that stage-wise dynamic voltage and frequency scaling (DVFS) is an effective optimization, allowing energy savings with only modest performance impact. Together, these findings offer practical insights and concrete guidance for designing more energy-efficient multimodal LLM serving systems.
\end{abstract}

%% file: sections/1-introduction.tex
\section{Introduction}

\begin{figure}
    \centering
    \includegraphics[width=0.7\linewidth]{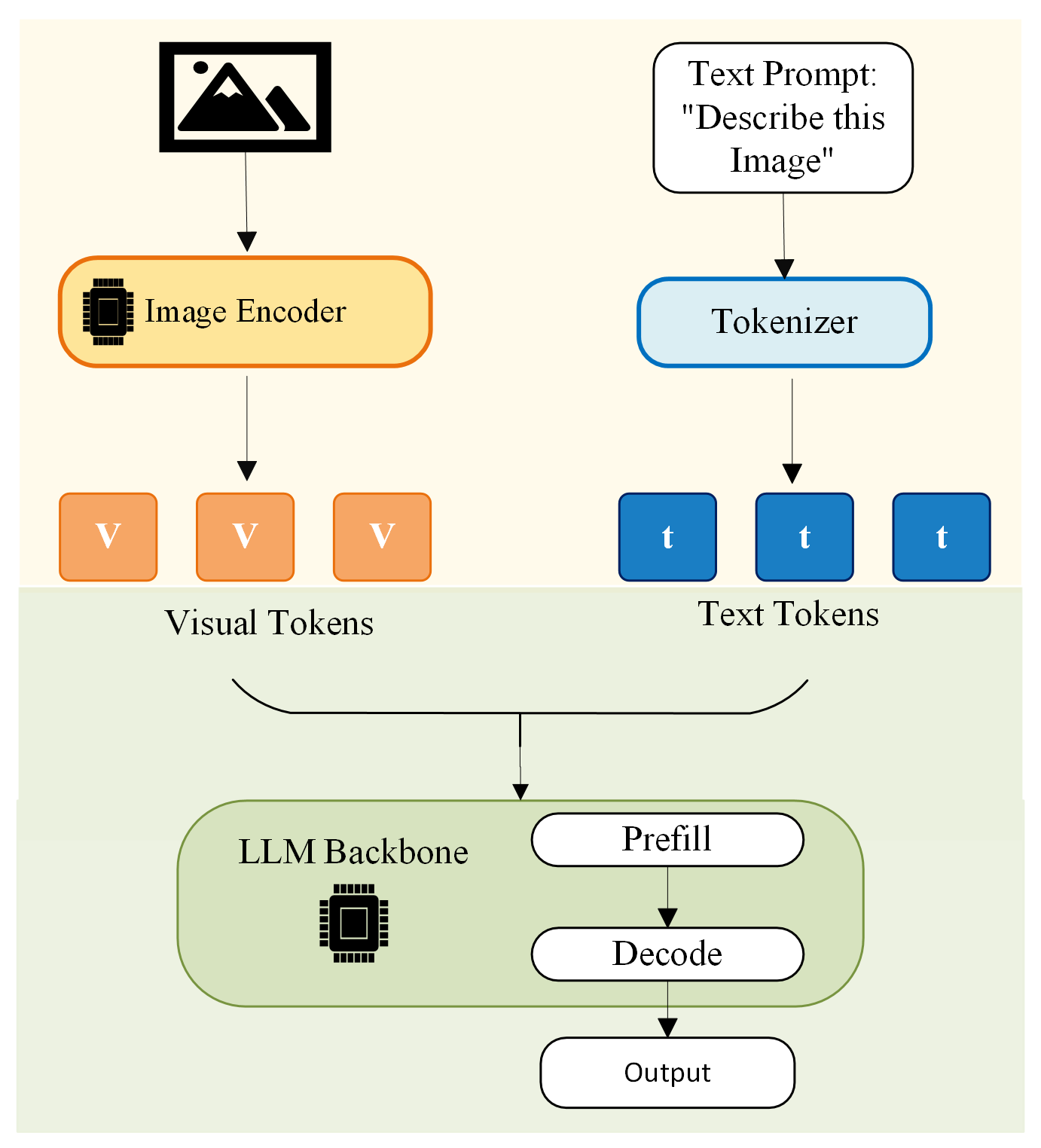}
    \caption{The MLLM inference pipeline. Multimodal inputs introduce a vision encoder and visual tokens (v), which are concatenated with text tokens (t) and inflate downstream prefill computation compared to text-only LLMs.}
    \label{fig:mllm-architecture}
\end{figure}

Large language models (LLMs) have undergone rapid advances in recent years, and their evolution has shifted toward multimodal large language models (MLLMs) \cite{wadekar2024evolution}. By integrating text with additional modalities such as images and videos, MLLMs extend the capabilities of LLMs and have become increasingly important for applications \cite{yin2024survey} including visual question answering, image captioning, and scene understanding \cite{schwenk2022okvqa,chen2022visualgpt,sarto2025image,fan2024mllm}. Despite their growing adoption, MLLMs differ from text-only LLMs in architectural complexity, workload characteristics, and resource demands. As illustrated in Figure \ref{fig:mllm-architecture}, MLLMs for image-text requests introduce a modality-specific encoder alongside the LLM backbone, which not only increases computational requirements but also produces hundreds to thousands of visual tokens that propagate into the prefill stage and reshape the inference pipeline \cite{chen2024image}. This additional stage substantially raises overall energy consumption, making energy efficiency a critical concern for large-scale deployment.

Meeting the escalating energy demands of MLLMs poses significant challenges for real-world serving systems. Prior work on text-only LLMs has shown that inference on modern GPUs is highly energy-intensive, with costs scaling superlinearly with model size and workload throughput \cite{niu2025tokenpowerbench}. These challenges are magnified in multimodal models, where vision encoders expand input complexity and inflate visual token counts \cite{cai2024vip}. Operating such energy-hungry workloads at datacenter scale raises sustainability concerns and operational costs, highlighting the need for energy-efficient serving strategies in large-scale MLLM deployments. To design energy-efficient systems, it is first necessary to develop a clear understanding of the models, workloads, and the sources of additional energy consumption for MLLMs.

Recent research has begun to address these challenges along two main directions. On the system side, several frameworks modularize MLLM architectures to enable more efficient serving and scheduling \cite{qiumodserve,singh2024efficiently,dong2025hydrainfer,liu2025elasticmm}. On the model side, techniques such as quantization, image compression, input sequence reduction, and token pruning reduce model size or limit the number of generated tokens, thereby lowering processing costs \cite{wang2024q,song2024less,shang2024llava}. However, while these approaches improve computation and throughput, they provide little insight into how energy consumption in MLLMs differs from text-only models or how it is distributed across inference stages, leaving opportunities for stage-wise energy optimization. This motivates our central objective to characterize how energy is distributed in multimodal pipelines and how input complexity propagates to higher energy cost.

In this work, we conduct a stage-level characterization of MLLM inference to understand where energy is consumed and how input complexity propagates to higher costs. Our study not only highlights architectural and workload-driven inefficiencies, but also evaluates a practical serving strategy using DVFS to improve energy efficiency.
Specifically, our contributions are:

\begin{itemize}
    \item A stage-level energy characterization of MLLM inference, quantifying energy distribution across encoder, prefill, and decoding stages relative to text-only LLMs.

    \item An analysis of how input complexity (resolution, image count, and visual tokens) affects energy scaling, revealing modality-inflation-driven inefficiencies through empirical power-trace analysis.

    \item An evaluation of stage-wise DVFS as a practical energy optimization strategy, deriving guidelines for efficient multimodal serving systems.
\end{itemize}

Finally, we outline promising directions for future work, proposing design principles toward energy-efficient serving platforms for multimodal LLMs.

%% file: sections/2-background.tex
\section{Background and Motivation} 

\subsection{MLLM Architecture and Inference Pipeline}
MLLMs employ modality-specific encoders to transform non-text inputs, such as images or videos, into representations compatible with the LLM backbone. These encoders
extract visual features and convert them into visual tokens that are aligned
with the language model embedding space \cite{yin2024survey}. The inference pipeline
typically consists of three stages: \textit{modality encoding}, \textit{prefill}, and
\textit{decoding}. In the encoding stage, the vision encoder generates visual
tokens; this stage generally exhibits computational and memory behavior that lies
between prefill and decoding \cite{dong2025hydrainfer}. The generated visual tokens are
then concatenated with text tokens and processed during the prefill stage, whose
cost scales with total sequence length and increases as visual tokens grow.
Finally, decoding generates output tokens autoregressively while attending to the
accumulating context window \cite{qiumodserve}. Compared to text-only inference, the
additional encoding stage and the inflated input sequence introduce extra
computation and memory overhead, fundamentally influencing the energy and
performance characteristics of MLLM inference.

\subsection{Encoder Architecture} 
In MLLMs, the vision encoder plays a central role in shaping inference cost by determining how visual inputs are represented and how many tokens are exposed to the language model. Unlike text-only inference, the output size of this encoder varies widely across architectures, making encoder design a key factor in sequence length, memory footprint, and computational load.
Existing MLLMs adopt different strategies for converting visual inputs into
tokens. A common approach is fixed-patch tokenization, used in ViT-based models
such as CLIP, where images are divided into uniform patches to produce a
constant token count \cite{radford2021learning,liu2024improved}. This yields predictable costs but does not scale with
higher resolutions. In contrast, any-resolution tiling divides images into
multiple crops or tiles, allowing flexible input handling at the cost of token
counts that grow with resolution and aspect ratio \cite{li2024llava,guo2024llava,luo2024feast}.
To mitigate token growth, some models compress visual features within the encoder
through downsampling or lightweight attention \cite{zhu2025internvl3,bai2025qwen2}, while query-based architectures
(e.g., Q-Former) project dense features into a fixed set of queries to bound
token counts \cite{li2023blip,kim2024towards}. Additional mechanisms such as resampling, condensation, and
token pruning further regulate token budgets \cite{alayrac2022flamingo,li2025tokenpacker,shang2024llava}.


\subsection{Energy Saving Controls for GPUs}
Modern GPUs expose several mechanisms to balance performance and power during
inference. A widely adopted control is \textit{Dynamic Voltage and Frequency
Scaling} (DVFS), which adjusts core or memory frequencies to reduce power
consumption with limited performance impact \cite{liu2025greenllm, maliakel2025investigating}. Recent
work shows that modest frequency reductions can yield noticeable energy savings
for LLM workloads \cite{10946802}.
Power capping offers another mechanism for managing GPU consumption by enforcing an upper power limit during execution. While effective for reducing peak power draw, the performance impact depends on workload characteristics and the severity of the cap
\cite{patel2024characterizing}. In this work, we focus on DVFS as a practical knob and evaluate its impact on
energy efficiency during MLLM inference.

\subsection{Modality Inflation Cost}
Although energy demand is already a growing concern for LLM inference \cite{kim2024towards,kakolyris2024slo}, multimodal workloads introduce an additional source of overhead by expanding the effective input sequence through visual tokens. This expansion increases both computation and memory traffic during inference and alters how energy is distributed across the pipeline. This modality-driven
growth, which we refer to as modality inflation, increases both compute
demand and memory traffic within the inference pipeline.

Despite its practical relevance, the energy implications of modality inflation
remain poorly understood. Existing literature predominantly optimizes throughput
and response latency, while the incremental energy cost introduced by the vision
encoder and inflated visual-token sequences has not been systematically
quantified. It is therefore unclear how much additional energy multimodal
inference consumes compared to text-only inference, or how different models and
inputs contribute to this growth.

\subsection{Workload Heterogeneity in MLLM}
\label{sec:workload-heterogeneity}
\begin{figure}[h!]
    \centering
    \begin{subfigure}{0.48\linewidth}
        \includegraphics[width=\linewidth]{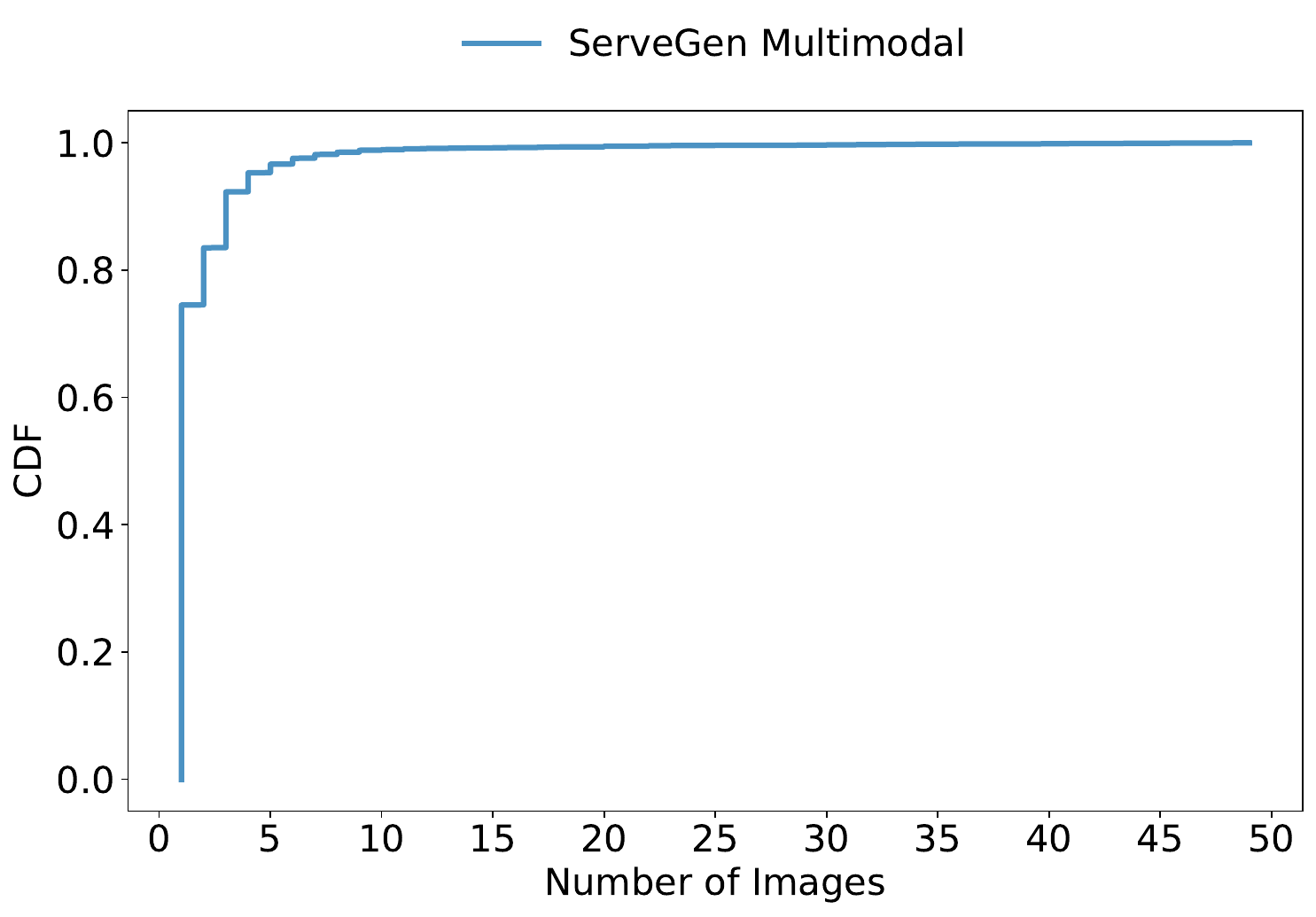} 
        \caption{Distribution of images per query from the ServeGen trace \cite{xiang2025servegen}.}
        \label{fig:workload-heterogeneity:a}
    \end{subfigure}
    \hfill
    \begin{subfigure}{0.48\linewidth}
        \includegraphics[width=\linewidth]{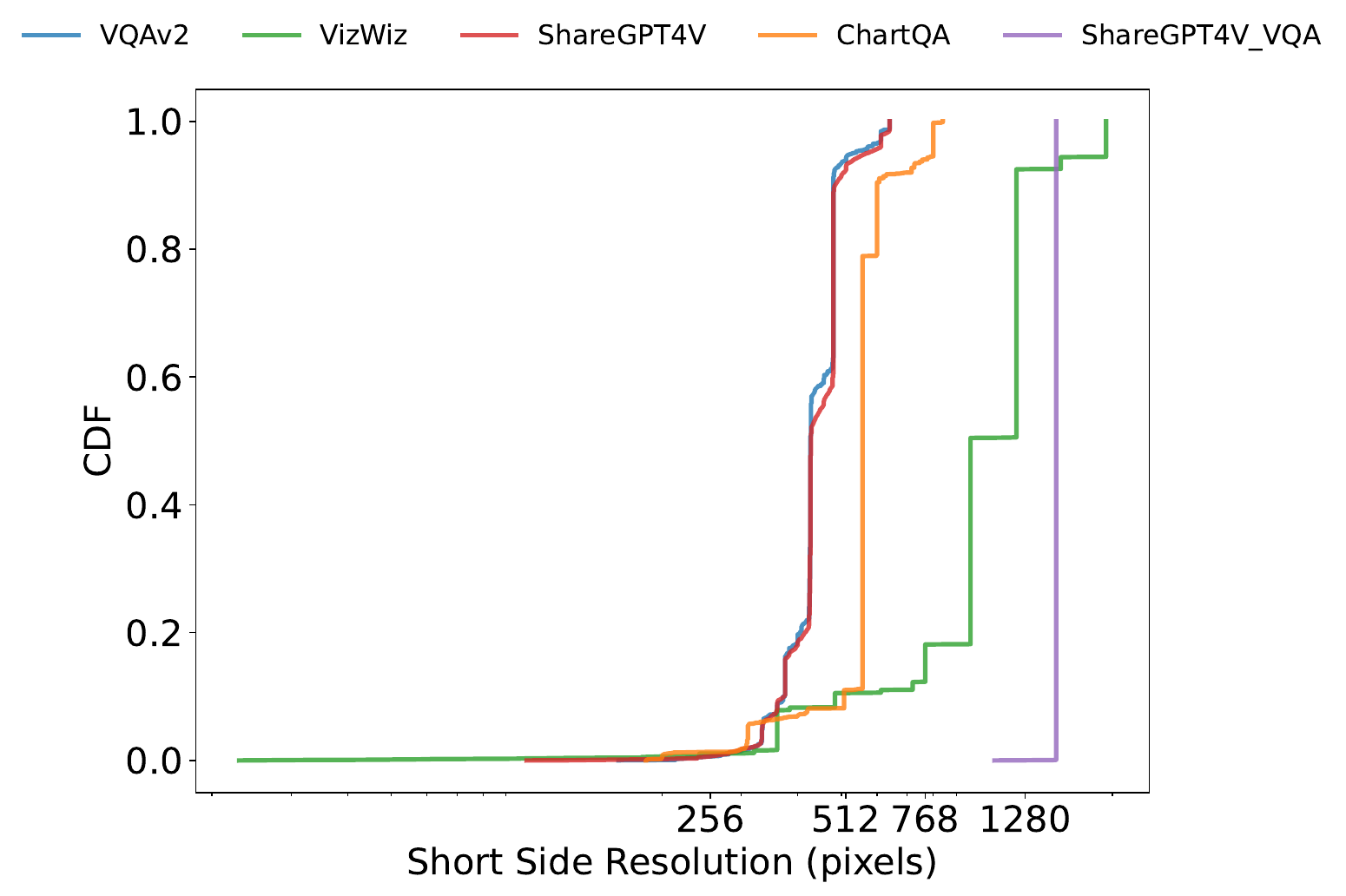} 
        \caption{Image resolution distributions across representative multimodal benchmark datasets.}
        \label{fig:workload-heterogeneity:b}
    \end{subfigure}
    \caption{Analysis of workload heterogeneity in multimodal requests.}
    
    \label{fig:workload-heterogeneity}
\end{figure}

There is a substantial degree of heterogeneity in multimodal requests. To better understand this behavior, we begin by analyzing the ServeGen \cite{xiang2025servegen} trace of production-like queries and showing the cumulative distribution function of the number of images per query. As shown in Figure \ref{fig:workload-heterogeneity:a}, many queries attach only one or two images. At the same time, we also observe queries with several dozen images, with rare extreme cases reaching up to 49 images per request. These image-heavy queries are relatively rare, but they still matter for the system. A small set of such requests is enough to slow down responses and increase resource usage, even though the average number of images per query is quite modest.
We then place these findings in context by looking at image resolution across four widely used benchmark datasets for multimodal tasks: VQAv2 \cite{goyal2017making}, VizWiz \cite{gurari2018vizwiz}, ShareGPT4V \cite{chen2024sharegpt4v}, and ChartQA \cite{masry2022chartqa}. Figure \ref{fig:workload-heterogeneity:b} shows the cumulative distribution functions of image resolution for these datasets. The images range from small, low-resolution inputs to very large images with millions of pixels. The four curves are not identical. Some datasets are skewed toward smaller images, while others contain a larger fraction of high-resolution content. As a result, two queries that include the same number of images can still require very different amounts of visual-token computation, depending on which dataset or application they come from.

Taken together, these observations show that multimodal workloads cannot be characterized using text tokens alone. Variability in image-related features, in particular the number of images and their resolution, plays a central role in “modality inflation,” where a minority of image-heavy, high-resolution requests disproportionately increase computation, energy consumption, and latency.

%% file: sections/4-characterizing.tex
\section{Characterizing Energy and Power in MLLM Inference}

\begin{table*}[t]
\centering
\renewcommand{\arraystretch}{1.15}

\begin{tabular}{%
p{0.28\textwidth}
p{0.25\textwidth}
p{0.25\textwidth}
p{0.12\textwidth}}
\hline
\textbf{Model} &
\textbf{Vision Encoder} &
\textbf{LLM Backbone} &
\textbf{Avg. Acc. (\%)} \\
\hline
InternVL3-8B \cite{zhu2025internvl3} &
InternViT-300M-v2.5 &
Qwen2.5-7B &
73.6 \\

LLaVA-1.5-7B \cite{liu2024improved} &
CLIP ViT-L/14 &
Vicuna-v1.5-7B &
36.9 \\

LLaVA-OneVision-Qwen2-7B \cite{li2024llava} &
SigLIP ViT &
Qwen2-7B &
60.2 \\

Qwen2.5-VL-7B \cite{bai2025qwen2} &
QwenViT &
Qwen2.5-7B &
70.9 \\
\hline
\end{tabular}

\caption{Architecture overview of the MLLMs evaluated in this study.}
\label{tab:models}
\end{table*}

In this section, we describe our experimental setup, including the hardware platform and model selection, and then present a stage-level characterization of MLLM inference and systematically decompose the inference pipeline to quantify the energy overhead introduced by visual modalities and analyze how multimodal inputs affect system behavior.
We investigate the following research questions:
\begin{itemize}
\item \textbf{RQ1:} What is the incremental energy cost of processing visual modalities relative to text-only inference on the same LLM backbone?\label{rq:modality-comparison}

\item \textbf{RQ2:} How is energy distributed across the inference pipeline, and to what extent does the vision encoder dominate total consumption?

\item \textbf{RQ3:} How does multimodal inference change the temporal GPU power profile relative to text-only inference, and what do the resulting phases imply for power-management opportunities?

\item \textbf{RQ4}: How do input complexity factors (e.g., resolution and image count) govern energy scaling across different encoder architectures?

\end{itemize}

\subsection{Experimental Setup}

\noindent\textbf{Models.} We select four representative MLLMs, including LLaVA-1.5 \cite{liu2024improved}, LLaVA-OneVision\cite{li2024llava}, Qwen2.5-VL\cite{bai2025qwen2}, and InternVL3 \cite{zhu2025internvl3}, which span a range of vision encoder designs and tokenization strategies, as explained in Table \ref{tab:models}. Together, these models represent common architectural choices in modern MLLMs, including fixed-patch tokenization, any-resolution tiling, compressed ViT-based encoding, and query-based token reduction. All models use LLM backbones in the 7B–8B parameter range, and three of them share the same LLM backbone architecture. This controlled selection allows us to isolate the impact of vision encoding and visual tokenization on energy consumption and keep the backbone scale comparable. 


\noindent\textbf{Workload.} To isolate the impact of visual modalities, we fix the text prompt across all
experiments. For image inputs, we use two configurations: (1) an iso-resolution
setting fixed at $512\times512$, which lies within the commonly used resolution
range observed in real workloads (Figure~\ref{fig:workload-heterogeneity:b})
; and (2) scaled inputs, where we vary image resolution and image count to characterize the sensitivity of each architecture to input complexity.

\noindent\textbf{Metrics.}
We evaluate system efficiency using energy per request (J) and end-to-end latency (s). We additionally analyze GPU power draw (W) to identify utilization phases and throughput (requests per second) to assess serving capacity.

\noindent\textbf{Hardware and Software.} All experiments are conducted on a single NVIDIA A100 80GB GPU. Model inference is executed using the PyTorch \cite{paszke2019pytorch} deep learning framework with Hugging Face Transformers \cite{wolf2020transformers}. To capture fine-grained energy profiles, we monitor GPU power consumption via the NVIDIA Management Library (NVML) \cite{nvidia_nvml}, sampling at 5 ms intervals. To ensure reproducibility, we allow utilization states to settle between consecutive runs and report averages over multiple repetitions with confidence intervals.






\subsection{Energy Overhead of MLLMs}
To address RQ1 and quantify the incremental energy cost of multimodal inference, we conduct a controlled iso-token comparison in which each MLLM is evaluated against a corresponding text-only baseline that uses the same LLM backbone. We match the total number of input tokens by equating the MLLM input length, including both text and visual tokens, to an equal number of text tokens in the baseline, and we fix the output to one token to remove variability from autoregressive decoding. This setup controls for sequence length effects and highlights the additional overhead introduced by multimodal processing.

\begin{figure}[h!]
    \centering
    \begin{subfigure}{0.48\linewidth}
        \includegraphics[width=\linewidth]{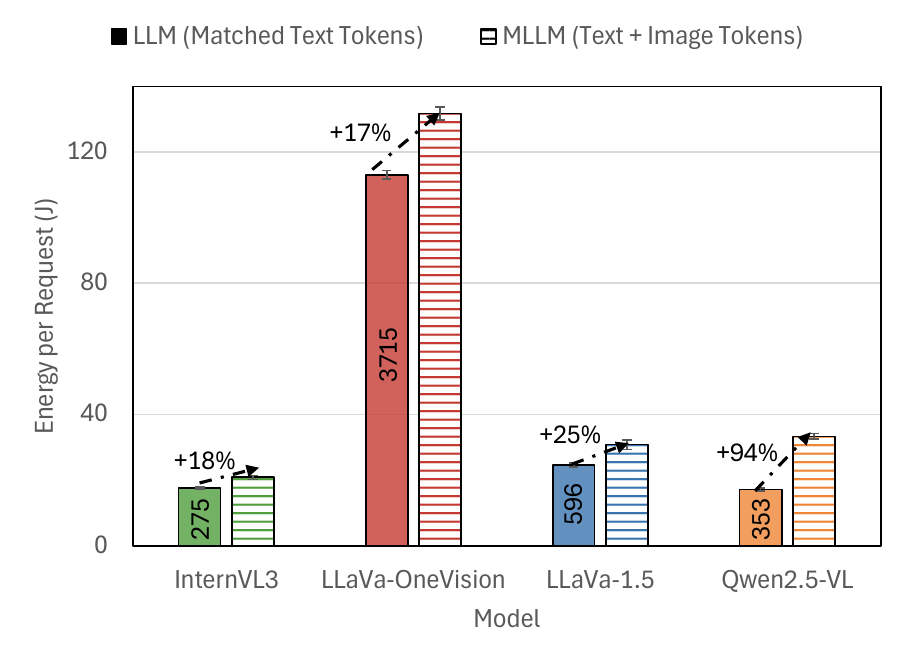} 
        \caption{Energy per Request}
    \end{subfigure}
    \hfill
    \begin{subfigure}{0.48\linewidth}
        \includegraphics[width=\linewidth]{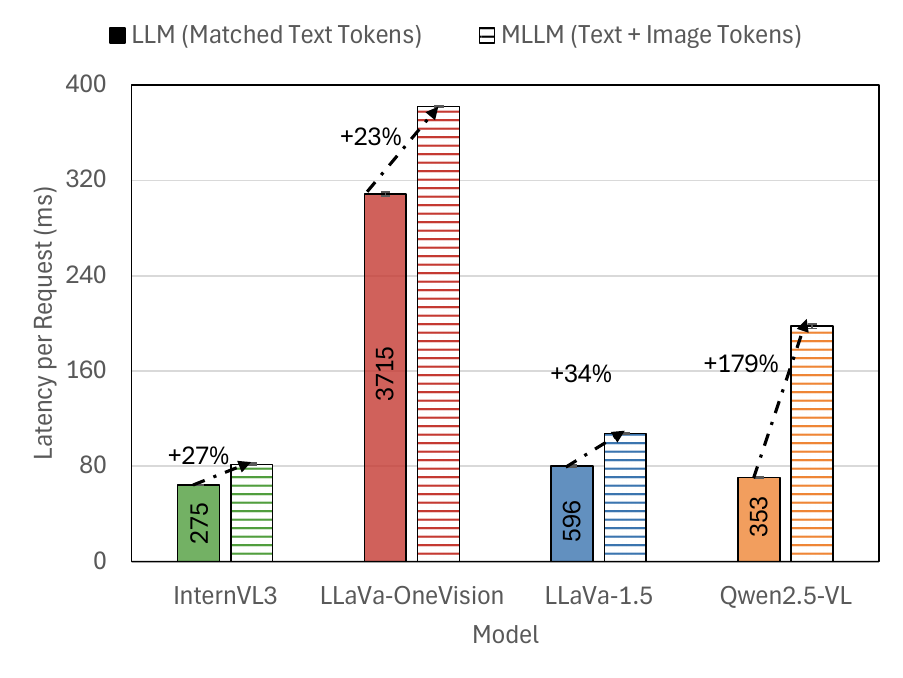} 
        \caption{Latency per Request}
    \end{subfigure}
    \caption{Iso-token comparison of (a) energy per request and (b) latency per request between multimodal models and their text-only baselines. The total input token count is matched by equating text-only tokens with the combined text and visual tokens in the multimodal setting, and output generation is fixed to one token. Numeric labels inside the LLM bars indicate the number of tokens used for iso-token matching. 
    }
    \label{fig:modality-cost}
\end{figure}

As shown in Figure~\ref{fig:modality-cost}, multimodal inference incurs a substantial energy overhead that varies widely across architectures, increasing energy per request by 17\% to 94\% relative to the matched text-only baseline. Qwen2.5-VL exhibits the largest overhead, with a 94\% energy increase, while LLaVA-1.5 and InternVL3 show more moderate increases of 25\% and 18\%, respectively. LLaVA-OneVision has the smallest relative energy increase at 17\%, despite producing 3{,}715 total input tokens, highlighting that token count alone does not determine energy overhead and that architectural choices in the multimodal front end play a major role.

Notably, energy and latency do not increase proportionally in this controlled setting. For Qwen2.5-VL, a 94\% energy increase corresponds to a 179\% latency increase, suggesting that a low parallelism stage can dominate end-to-end time. Overall, these results indicate that, unlike text-only LLM inference where energy behavior is relatively consistent across similarly sized models \cite{pronk2025benchmarking}, MLLMs can exhibit large architecture dependent differences even under controlled input conditions, motivating model specific serving policies rather than one-size-fits-all configurations. 


\observation{1}{The extreme variance in energy overhead across MLLMs motivates architecture-specific system configurations for energy efficiency, as generic policies cannot accommodate such wide behavioral differences.
}





\subsection{Stage-wise Characterization}
\begin{figure}[h!]
    \centering
    \begin{subfigure}{0.48\linewidth}
        \includegraphics[width=\linewidth]{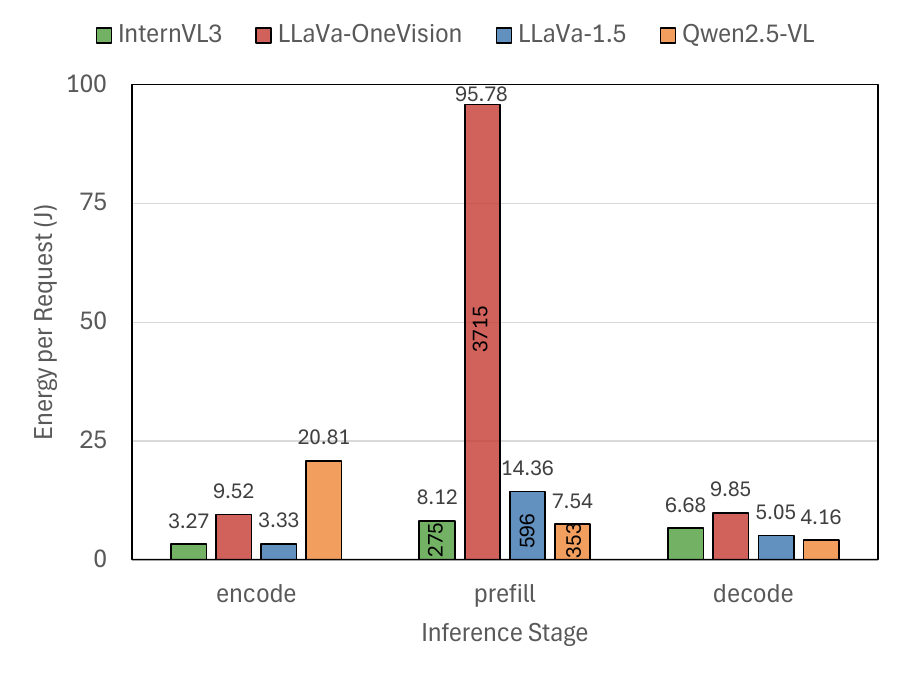} 
        \caption{Energy per Request}
    \end{subfigure}
    \hfill
    \begin{subfigure}{0.48\linewidth}
        \includegraphics[width=\linewidth]{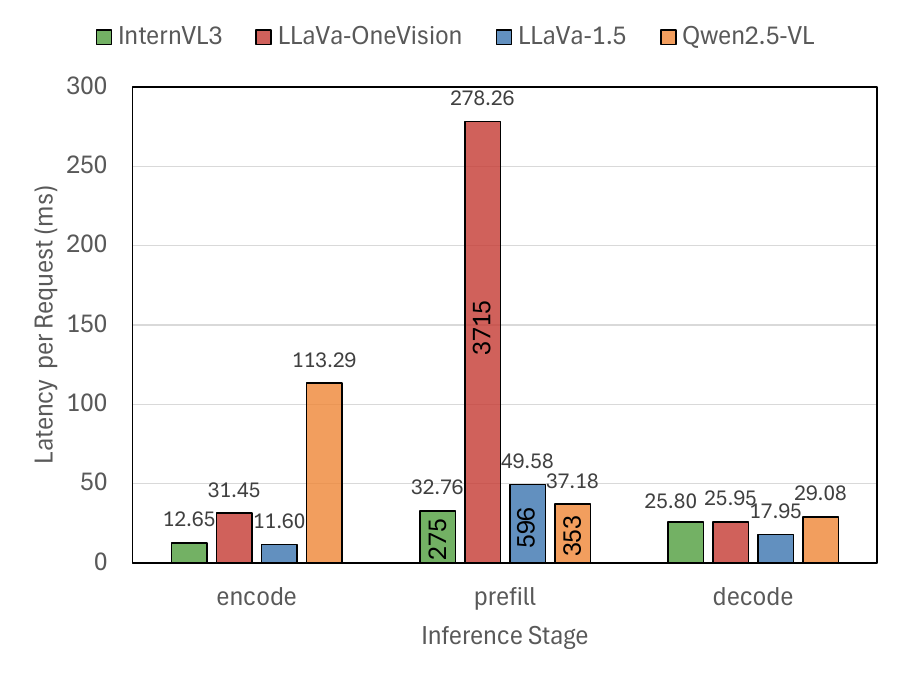} 
        \caption{Latency per Request}
    \end{subfigure}
    \caption{Stage-wise breakdown of inference into encoding, prefill, and decoding with a fixed output length, showing (a) energy per request and (b) latency per request for each stage. Numeric labels inside the prefill bars indicate the number of visual tokens produced by each model in this setting.
}
    \label{fig:stage-wise}
\end{figure}

While in the previous section we discussed that multimodal inputs incur a substantial energy penalty, treating the model as a black box hides the specific architectural sources of this inefficiency. To determine where this energy is consumed and to address RQ2, we break down the inference pipeline into vision encoding, prefill, and decoding stages, using a fixed output length of 32 tokens to capture decoding behavior consistently across models. Our analysis shows that the vision encoder influences system energy efficiency in two ways: a \emph{direct} cost from visual feature extraction that stems from its computational complexity, and an \emph{indirect} cost that propagates to the LLM prefill stage when visual inputs expand into long token sequences. As illustrated in Figure~\ref{fig:stage-wise}, these mechanisms create different bottlenecks across architectures.

Qwen2.5-VL exemplifies the impact of high direct overhead; its compute-intensive encoder consumes 20.81~J, which is $6\times$ higher than that of LLaVA-1.5. Additionally, an increase in latency of 113.29~ms indicates that this overhead substantially affects end-to-end performance. In contrast, although the encoder of LLaVA-OneVision is relatively efficient (9.52~J), its tiling strategy generates a large sequence of 3{,}715 visual tokens. This token expansion escalates the prefill stage energy to 95.78~J and increases prefill latency to 278.26~ms, about $12\times$ the energy and $8.5\times$ the latency of the balanced InternVL3 baseline (8.12~J, 32.76~ms).

The latency breakdown also highlights different optimization opportunities: for LLaVA-1.5, the encoding stage is short (about 12~ms), suggesting potential latency slack for energy-saving techniques, whereas for LLaVA-OneVision the end-to-end latency is dominated by prefill, making this stage the primary target under tight response-time constraints. Across all models, decoding remains comparatively stable, with energy largely governed by output length rather than encoder design or input length \cite{fernandez2025energy}. These results show that energy bottlenecks vary across stages depending on the architecture, motivating stage-specific configurations and stage-wise DVFS.

\observation{2}{Energy bottlenecks shift across stages depending on model architecture, motivating stage-specific configurations and stage-wise DVFS to maximize energy savings.}

\subsection{Power Profiling}
\begin{figure*}[t]
    \centering
    \begin{subfigure}{0.24\textwidth}
        \centering
        \includegraphics[width=\linewidth]{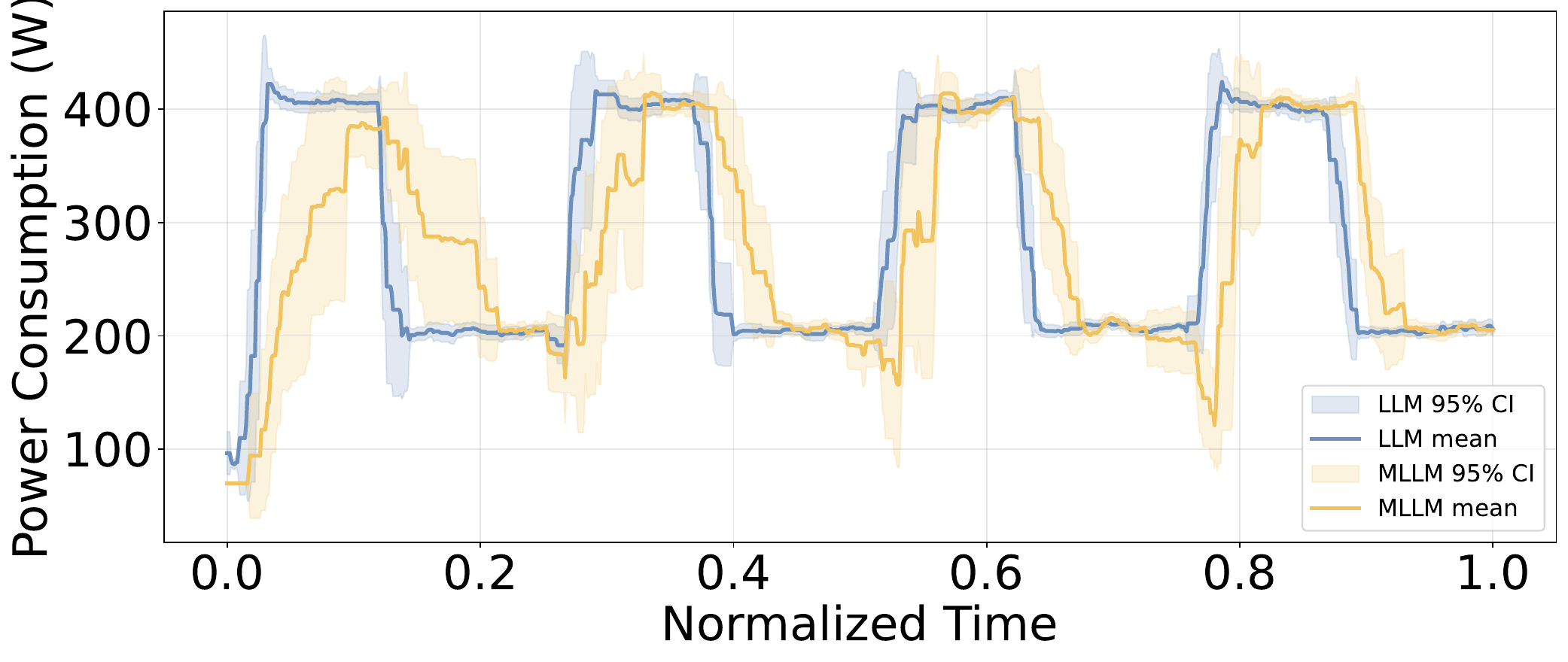}
        \caption{InternVL3}
    \end{subfigure}\hfill
    \begin{subfigure}{0.24\textwidth}
        \centering
        \includegraphics[width=\linewidth]{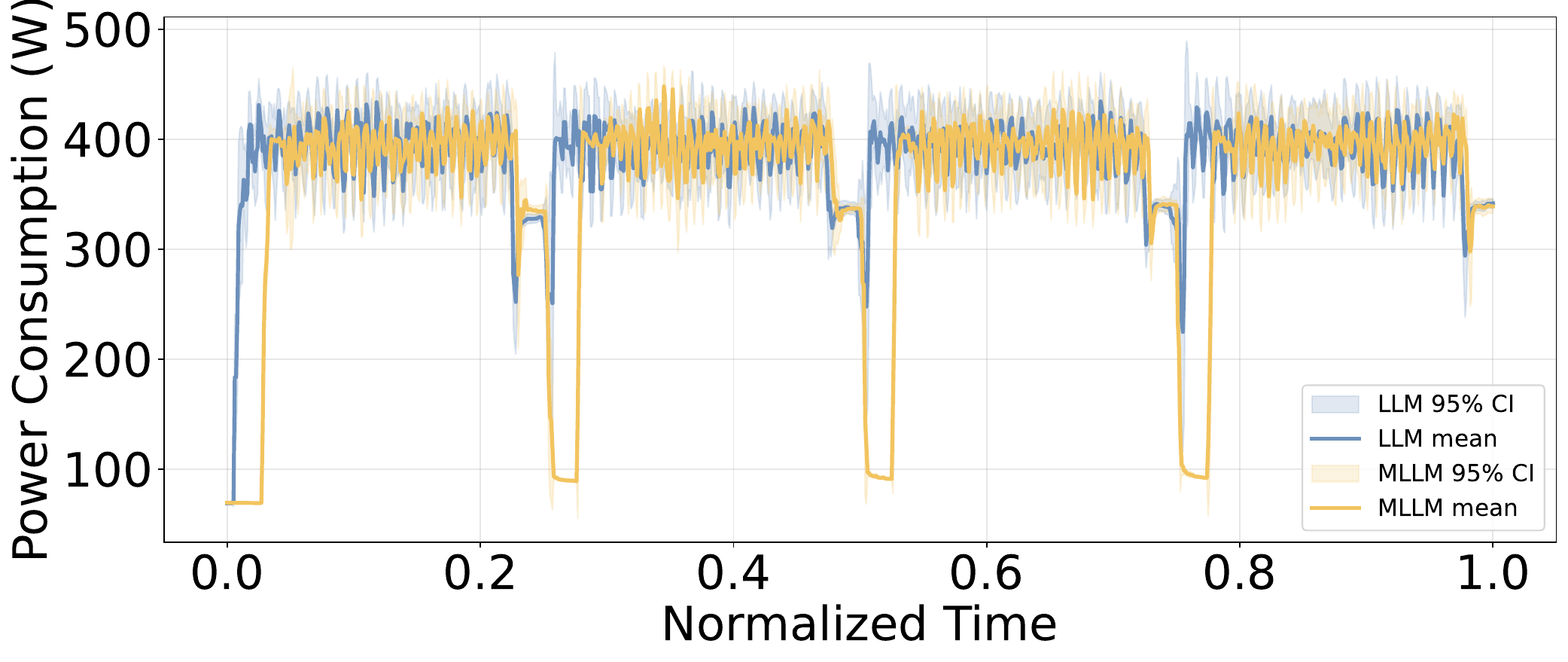}
        \caption{LLaVA-OneVision}
    \end{subfigure}\hfill
    \begin{subfigure}{0.24\textwidth}
        \centering
        \includegraphics[width=\linewidth]{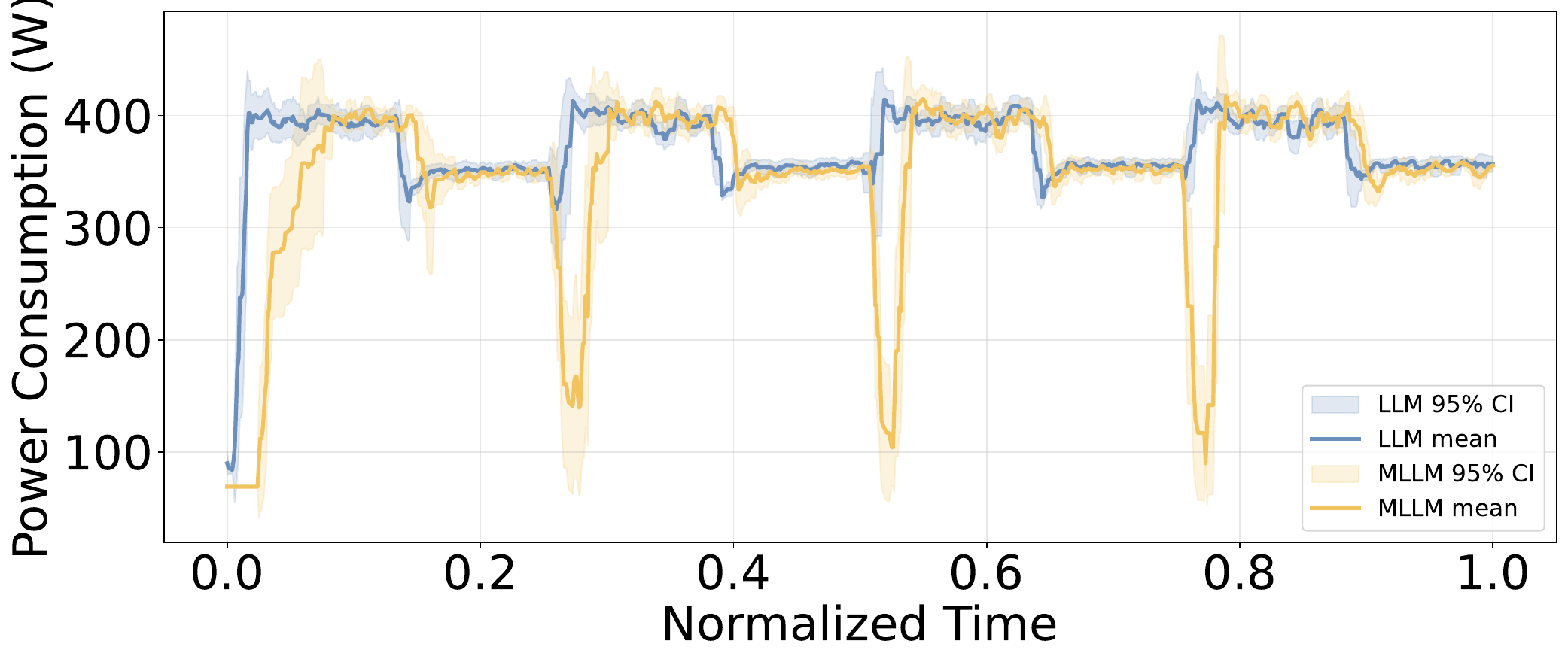}
        \caption{LLaVA-1.5}
    \end{subfigure}\hfill
    \begin{subfigure}{0.24\textwidth}
        \centering
        \includegraphics[width=\linewidth]{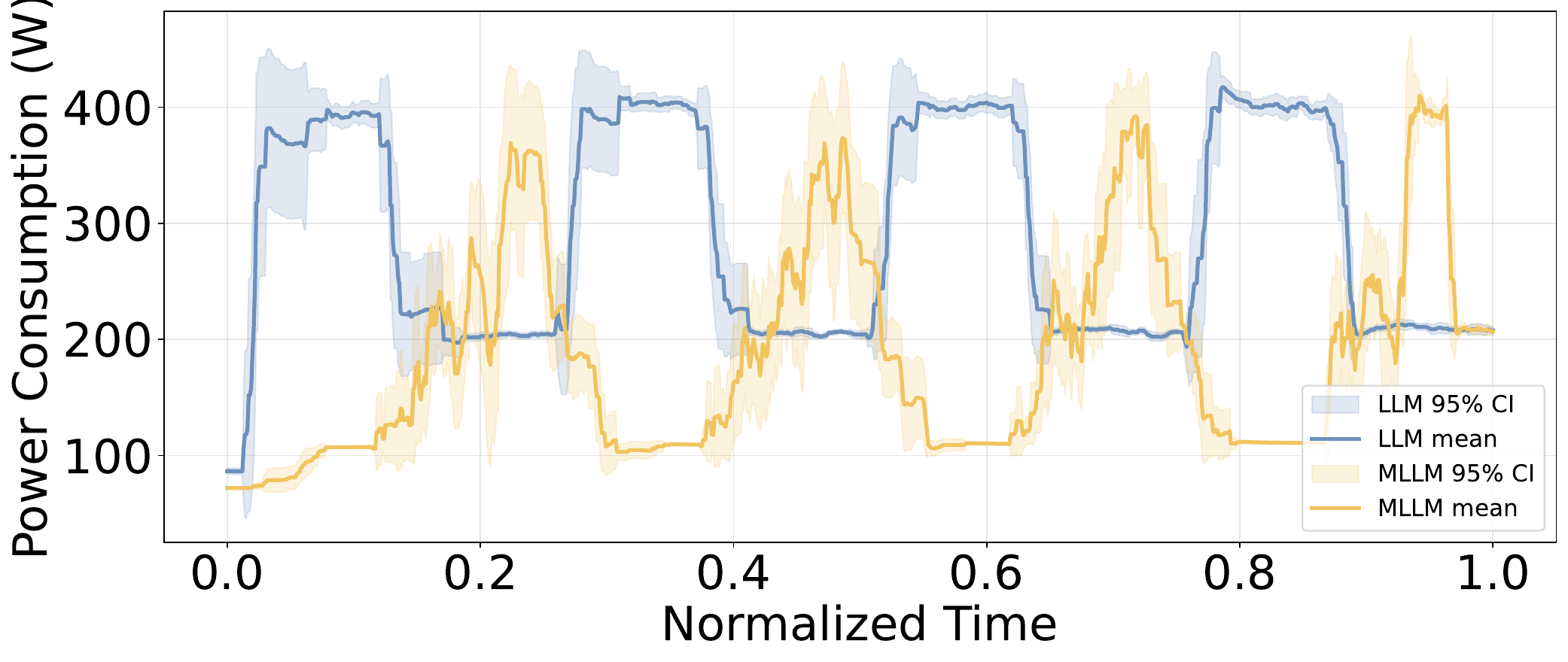}
        \caption{Qwen2.5-VL}
    \end{subfigure}

    \caption{ Normalized GPU power traces for text-only LLM and multimodal LLM inference using batch size 32 and output length 32 to drive the GPU toward saturation.}
    \label{fig:power-comparison}
\end{figure*}
To answer RQ3, we analyze instantaneous GPU power draw during inference by comparing normalized power traces of multimodal requests to iso-token text-only baselines. For these experiments, we use a batch size of 32 to represent a throughput optimized setting that drives the A100 toward saturation. As shown in Figure~\ref{fig:power-comparison}, the text-only baseline transitions rapidly from idle (about 80~W) to near the GPU power limit (about 400~W), indicating sustained high GPU activity in this optimized setting. In contrast, multimodal inference exhibits extended mid power phases, typically between 100~W and 250~W, which coincide with the vision encoding portion of the pipeline. Qwen2.5-VL shows the most pronounced effect, with a step-like power pattern around 200~W for a significant duration before the LLM backbone engages, while LLaVA-OneVision exhibits higher frequency fluctuations consistent with bursty tile processing. Importantly, while disaggregated serving can increase encoder batch size and improve utilization \cite{qiumodserve,dong2025hydrainfer}, such batching is constrained by stochastic request arrivals and end-to-end latency targets; therefore, mid power phases can still arise under realistic operating points. Overall, these trace signatures indicate that multimodal execution spends substantial time in a lower power regime than the saturated text-only baseline, suggesting that generic race-to-idle frequency policies may keep the GPU at high clocks during phases that do not benefit proportionally from higher frequency. 

\observation{3}{Multimodal inference introduces mid-level power phases in vision encoding that are not observed in text-only inference. These phases illustrate how the GPU remains in a lower power mode for an extended period, which is less than in fully saturated text inference, suggesting that default frequency scaling rules may not be suitable for multimodal execution.}
\begin{figure}[h!]
    \centering
    \begin{subfigure}{0.48\linewidth}
        \includegraphics[width=\linewidth]{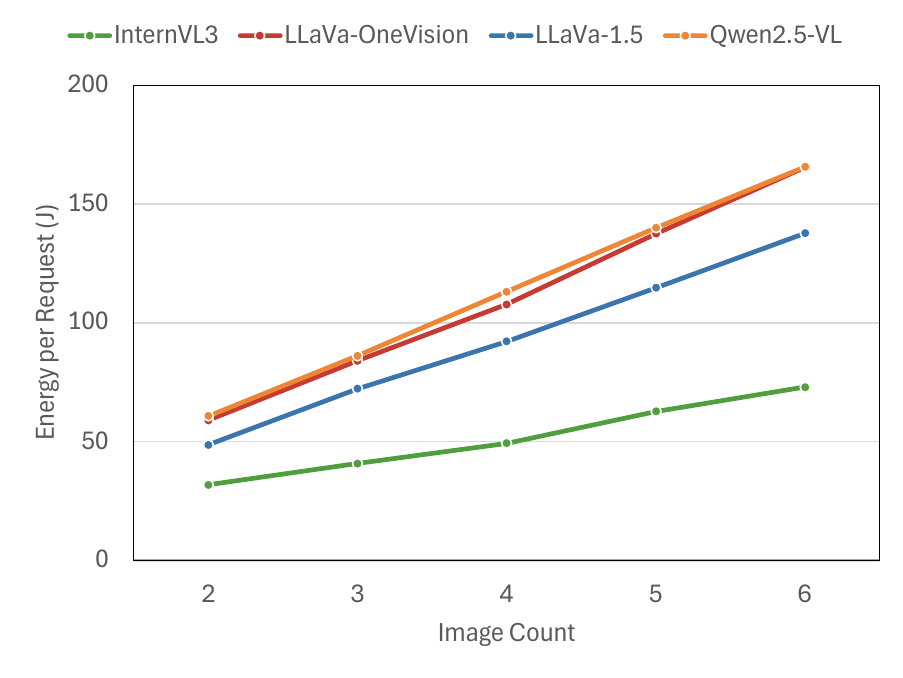}
      
        \caption{Energy per Request}
    \end{subfigure}
    \hfill
    \begin{subfigure}{0.48\linewidth}
        \includegraphics[width=\linewidth]{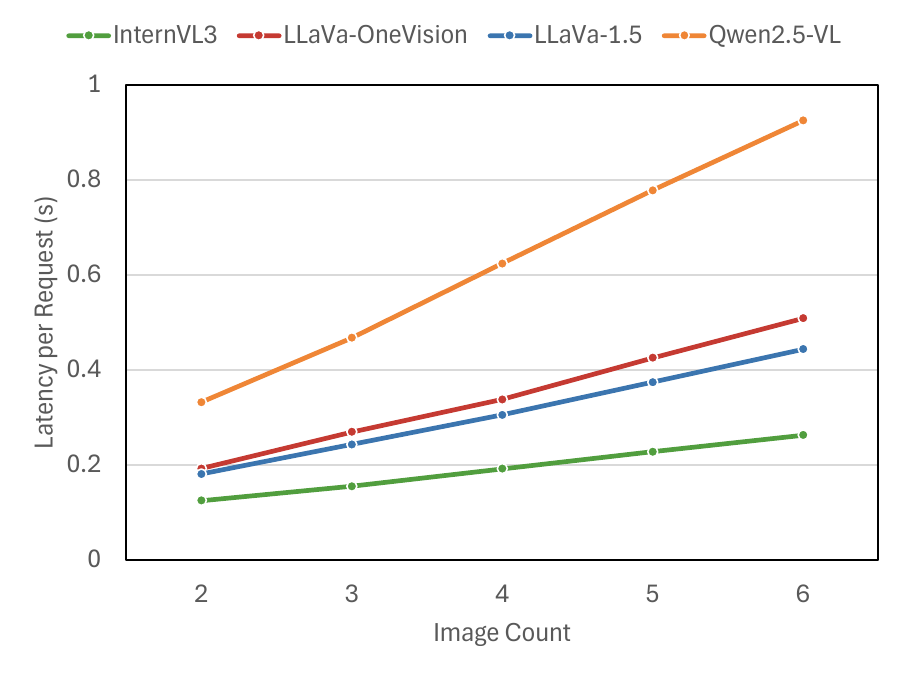} 
        
        \caption{Latency per Request}
    \end{subfigure}
    \caption{Impact of image count on energy and latency in MLLM inference.}
    \label{fig:inputComplexity-imageCount}
\end{figure}
\subsection{Input Complexity}
\begin{figure*}[t]
    \centering
    \begin{subfigure}{0.32\linewidth}
        \includegraphics[width=\linewidth]{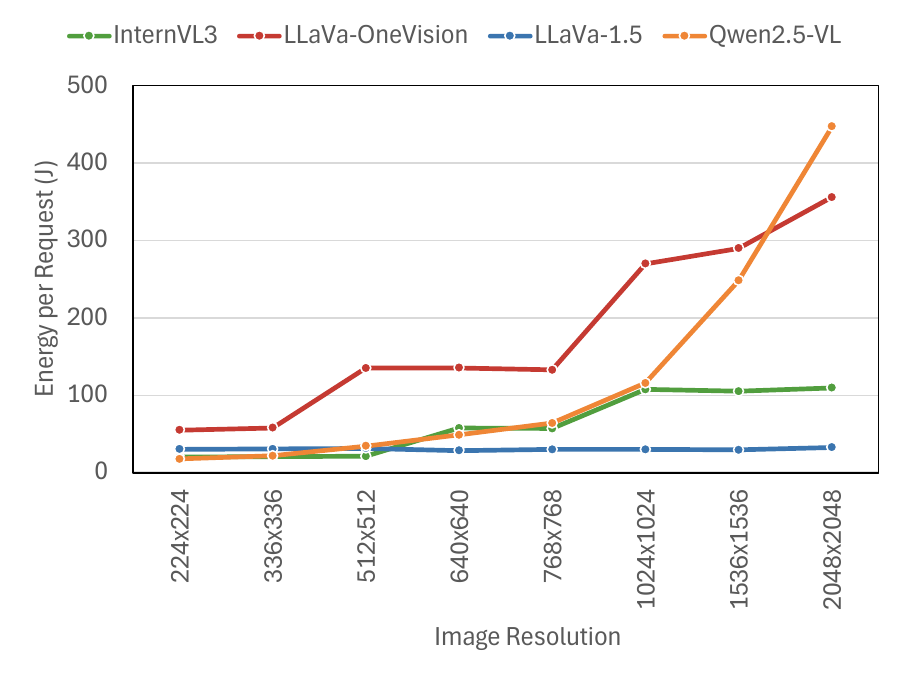}
        \caption{Energy per Request}
            \label{fig:imageResolution-scaling-energy:a}
    \end{subfigure}
    \hfill
    \begin{subfigure}{0.32\linewidth}
        \includegraphics[width=\linewidth]{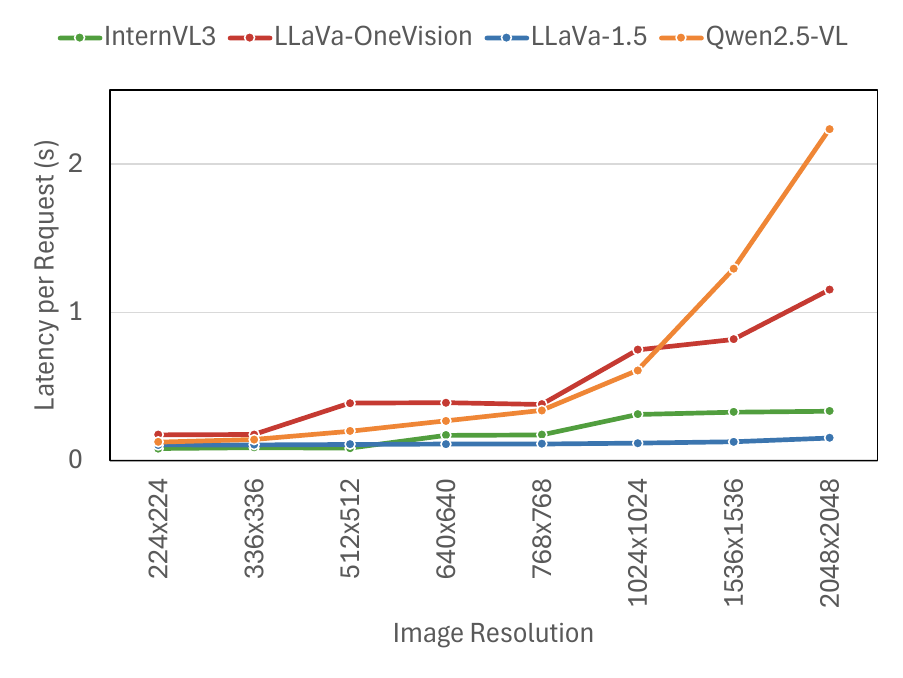}
        \caption{Latency per Request}
        \label{fig:imageResolution-scaling-latency:b}
    \end{subfigure}
    \hfill
    \begin{subfigure}{0.32\linewidth}
        \includegraphics[width=\linewidth]{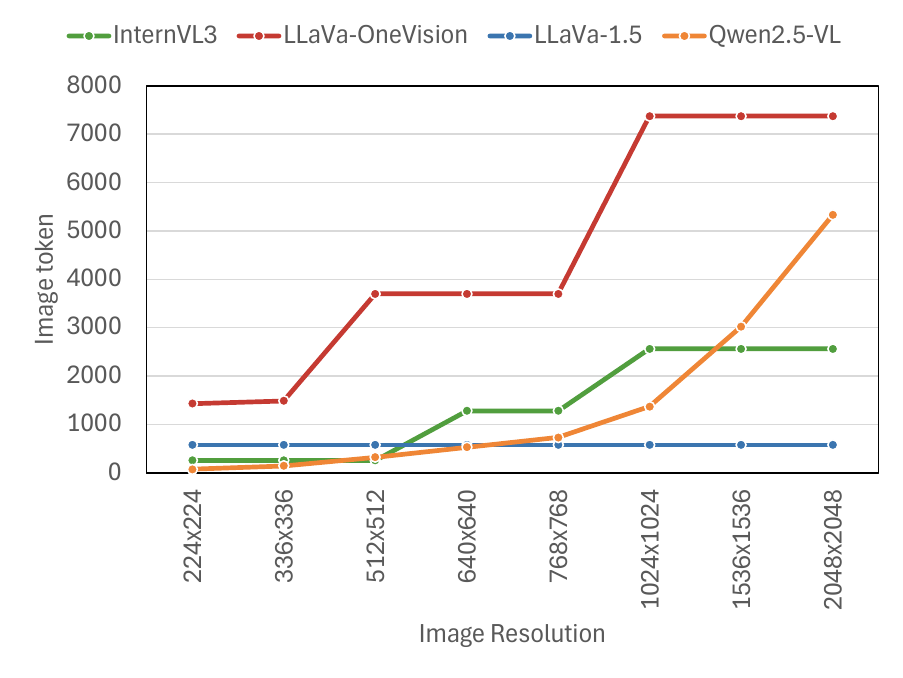}
        \caption{Visual Tokens vs Resolution}
        \label{fig:imageResolution-scaling-token:c}
    \end{subfigure}

    \caption{Impact of image resolution on energy, latency, and visual token generation in multimodal LLM inference. }
    \label{fig:imageResolution-scaling}
\end{figure*}

To address RQ4, we characterize how energy and latency scale with two primary dimensions of multimodal input complexity, image resolution and image count, while keeping the text prompt fixed. This analysis is important for serving because real workloads vary widely in both resolution and the number of images per request, as discussed in Section~\ref{sec:workload-heterogeneity}, and worst-case inputs can dominate both energy consumption and tail latency. As shown in Figure~\ref{fig:imageResolution-scaling-energy:a} and \ref{fig:imageResolution-scaling-latency:b}, both energy and latency exhibit clearly different scaling behaviors across architectures. LLaVA-1.5 remains nearly flat as resolution increases, consistent with its fixed-resolution CLIP encoder that resizes inputs before tokenization. InternVL3 and LLaVA-OneVision exhibit non-uniform scaling over the evaluated resolution range, with noticeable increases at specific resolution settings. These trends are explained by how each model expands images into visual tokens. Figure~\ref{fig:imageResolution-scaling-token:c} shows that LLaVA-1.5 produces an almost constant number of visual tokens across resolutions, whereas InternVL3 and LLaVA-OneVision increase token counts discretely over the tested resolutions, and Qwen2.5-VL exhibits rapid token growth at higher resolutions. In particular, Qwen2.5-VL shows a sharp increase in both energy and latency beyond $1024\times1024$, and the accompanying token expansion suggests that high-resolution inputs can substantially increase prefill workload when token growth is not effectively controlled. Beyond resolution, Figure \ref{fig:inputComplexity-imageCount} shows that increasing the number of images increases both energy and latency over the evaluated range, but with markedly different slopes across models. These slopes capture the marginal cost per additional image and directly constrain multi-image serving under fixed response-time targets.

\observation{4}{Multimodal architectures exhibit fundamentally different sensitivities to input complexity. Consequently, achieving system efficiency necessitates input-aware techniques that account for these distinct scaling behaviors.}

%% file: sections/5-controls.tex
\section{Discussion: Stage-wise GPU Frequency Scaling}

Our experiments demonstrate that the energy consumption and efficiency of MLLM inference vary widely based on the characteristics of the inputs, the vision encoder structure, and the extent of visual token expansion. This suggests that naive, one-size-fits-all serving policies for MLLM inference lead to inefficient use of hardware resources, highlighting the need for energy-efficient and MLLM-specific serving mechanisms. We select InternVL3 and Qwen2.5-VL as case studies because they represent two contrasting energy bottlenecks in multimodal inference. Qwen2.5-VL is dominated by a compute-intensive vision encoder, whereas InternVL3 exhibits lower encoding cost but substantial prefill energy overhead. Hence, in this discussion, our main focus is on the vision encoding and prefill energy components that dominate the energy overhead of multimodal inputs. We do not consider the energy of the decoding components, as the related energy has been thoroughly studied in the text-based LLM with scalability to the length of the outputs~\cite{liu2025greenllm}.
From both models, the energy per request heatmaps in each stage in Figure \ref{fig:dvfs} demonstrate that a single global serving method where the GPU runs at its highest frequency does not necessarily result in minimum energy consumption per request during multimodal inference tasks. In these experiments, the energy heatmaps show that in most stages, the energy per request has a tendency to be at its minimum at intermediate frequencies rather than at the highest clock rate, which distinguishes latency optimal points from energy optimal points.
Starting with InternVL3, the encode-stage heatmaps clearly indicate that significant opportunities for energy savings exist with limited performance impact. At a batch size of 32, increasing the GPU frequency from 1050 MHz to 1410 MHz reduces encoding latency from 0.18 s to 0.16 s (an 11.8\% reduction), while throughput improves by 13.4\%; however, energy per request increases from 1.03 J to 1.28 J (a 24.9\% increase). The encode energy heatmaps further show that energy remains close to its minimum at intermediate frequencies, even as latency continues to decrease at higher clocks. This behavior indicates diminishing latency returns for encoding at higher frequencies and suggests that adjusting the encoding frequency offers meaningful opportunities for energy savings with modest performance loss, particularly when downstream stages dominate end-to-end latency.

In contrast, despite its lightweight encoder, the prefill stage of InternVL3 accounts for the majority of the multimodal overhead once visual tokens are expanded. As shown in the prefill energy heatmaps, at a batch size of 32 and a frequency of 1050 MHz, prefill energy is 5.53 J compared to 1.03 J for encoding, accounting for approximately 84.3\% of the combined encode and prefill energy. Although prefill latency is more sensitive to frequency scaling than encoding latency, its energy minimum does not occur at the highest frequency. Increasing frequency from 1050 MHz to 1410 MHz reduces prefill latency from 0.72 s to 0.66 s (an 8.8\% reduction), while energy increases from 5.53 J to 6.12 J (a 10.6\% increase). The corresponding throughput heatmaps show continuous improvements with increasing frequency, albeit at the cost of higher energy consumption. These results indicate that, for InternVL3, the encoding stage presents substantial opportunities for downclocking, whereas the prefill stage should be tuned primarily to satisfy latency or throughput constraints rather than energy minimization.

Similar trends are observed for Qwen2.5-VL, where higher GPU frequencies reduce latency and improve throughput for both encoding and prefill stages at the cost of increased energy per request. However, in Qwen2.5-VL, the encoder dominates both end-to-end latency and a large fraction of energy consumption. In the prefill stage, the same qualitative behavior persists but with smaller absolute costs, as reflected in the prefill heatmaps. Increasing frequency from 1050 MHz to 1410 MHz reduces prefill latency from 0.88 s to 0.79 s (a 10.8\% reduction), while energy increases from 6.30 J to 7.40 J (a 16.5\% increase). Although prefill contributes less to overall latency, its energy behavior further confirms that no single frequency simultaneously optimizes latency, throughput, and energy.

Overall, these results demonstrate that stage-level DVFS is applicable to Qwen2.5-VL, but that the optimal configuration depends strongly on the serving objective. When responsiveness is critical—particularly for encoder-dominated models such as Qwen2.5-VL—higher frequencies may be justified despite increased energy consumption. Conversely, when latency slack exists, downclocking the dominant stage can significantly reduce energy per request at the cost of increased latency. These findings motivate an SLO-aware (service-level objective–aware) frequency selection mechanism rather than a fixed global setting. The design of such a frequency controller is left for future work.

%% file: sections/7-relatedWork.tex
\section{Related Work}
\subsection{Energy Optimization for LLMs}
Existing energy optimization techniques for large language models focus on text-only workloads and primarily propose optimizations for the prefill and decode stages of inference. These works explore a range of techniques, such as scheduling, batching, and power management, to reduce inference cost while maintaining acceptable latency \cite{liu2025greenllm, 10946802, fernandez2025energy, wilkins2024offline}. However, these approaches do not consider multimodal models and requests, where additional input modalities and encoder can significantly alter inference-stage behavior and introduce new sources of energy inefficiency. Unlike prior energy characterization studies that focus on text-only LLMs or earlier multimodal models under different hardware and workload assumptions, our work emphasizes controlled, within-study comparisons across modern vision–language models. 

\subsection{Efficient Serving Systems for MLLMs}
Current serving methodologies for MLLMs are mainly focused on improved throughput and lower end-to-end latency using techniques such as efficient batching, pipelined parallelism, or hardware/model disaggregation \cite{qiumodserve, liu2025elasticmm, dong2025hydrainfer, singh2024efficiently, guo2025rserve}. While these methodologies have achieved great performance improvements, they do not focus on energy efficiency. As a result, existing methodologies have not attempted to analyze energy tradeoffs among different elements in multimodal inference, including a vision encoder and a language model backbone, nor have they explored energy-efficient hardware control for multimodal inference.

\subsection{Power Management and DVFS in GPU-Based Systems}
Power management in both GPU-based systems and datacenter settings has been studied widely, with a focus on improving power efficiency under performance requirements. In existing work, a variety of approaches have been considered for managing power usage, such as DVFS, power capping, and hardware transformations, which have common usage in dealing with latency or throughput requirements \cite{patel2024characterizing, maliakel2025investigating}. The above approaches generally rely on characteristics of workloads, in addition to system state information, for making decisions on power management in order to lower power consumption without overstepping performance requirements.

\begin{figure}[t]
\centering
\small
\begin{subfigure}{0.30\linewidth}
    \includegraphics[width=\linewidth]{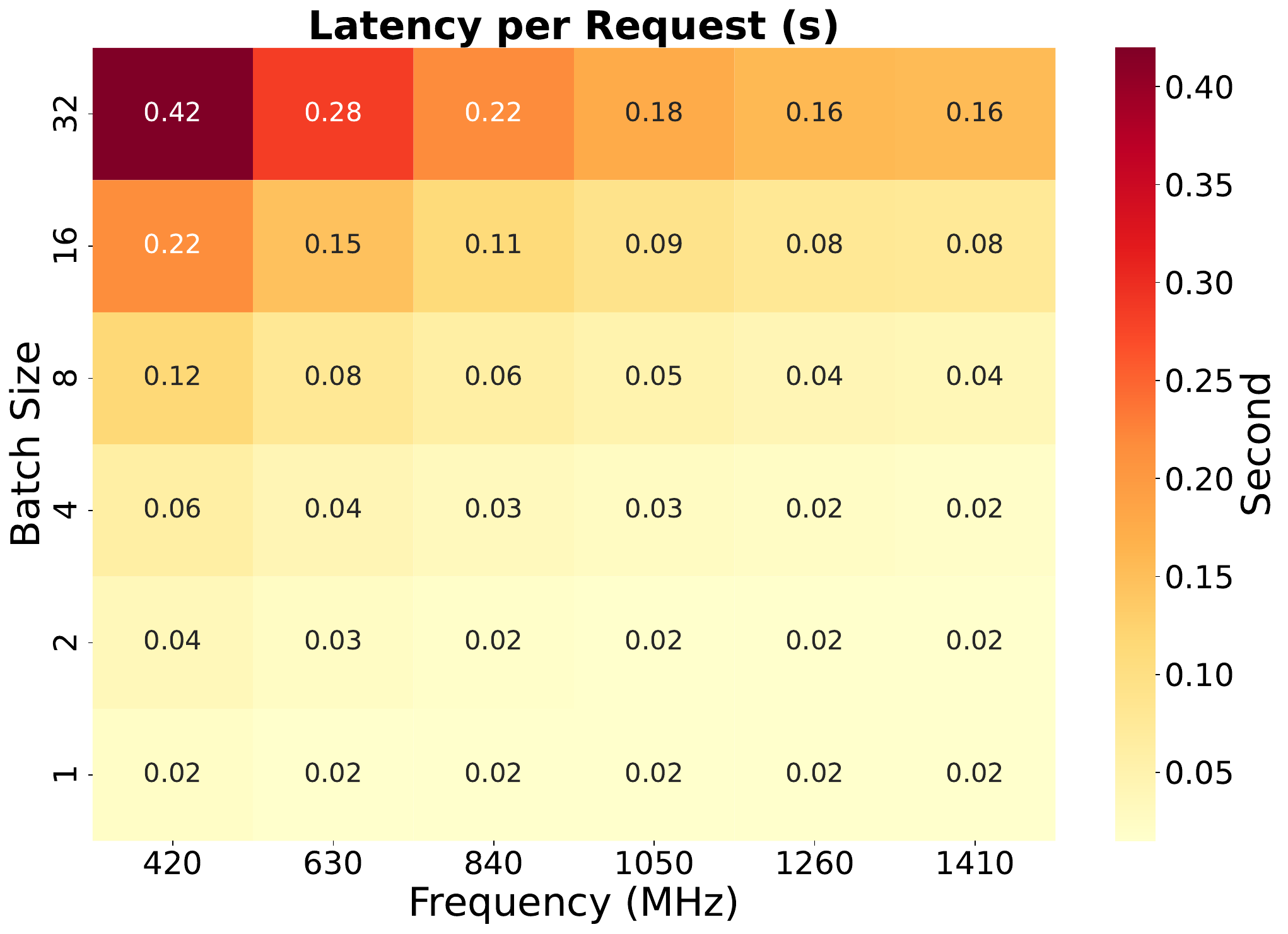}
\end{subfigure}\hfill
\begin{subfigure}{0.30\linewidth}
    \includegraphics[width=\linewidth]{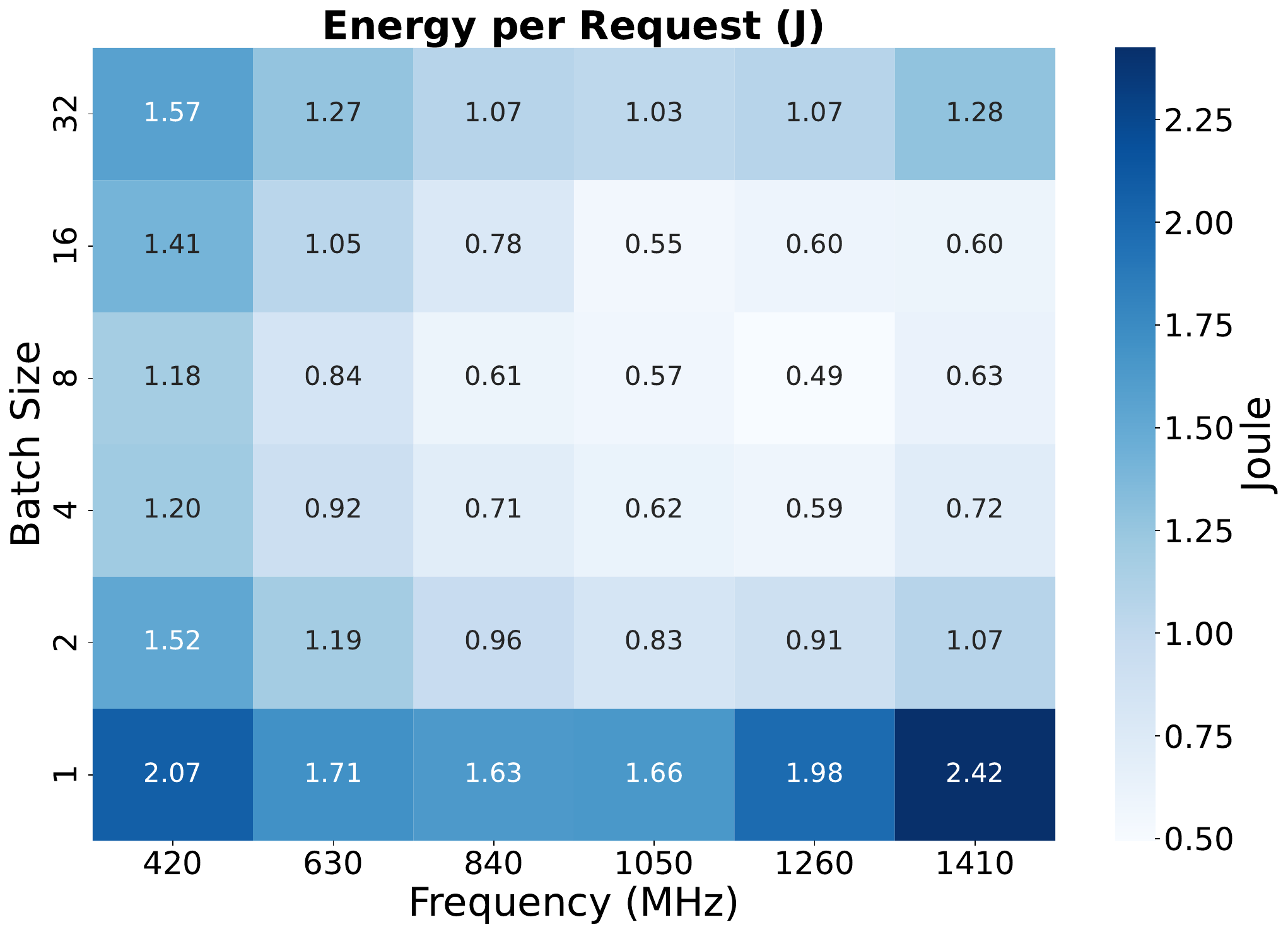}
\end{subfigure}\hfill
\begin{subfigure}{0.30\linewidth}
    \includegraphics[width=\linewidth]{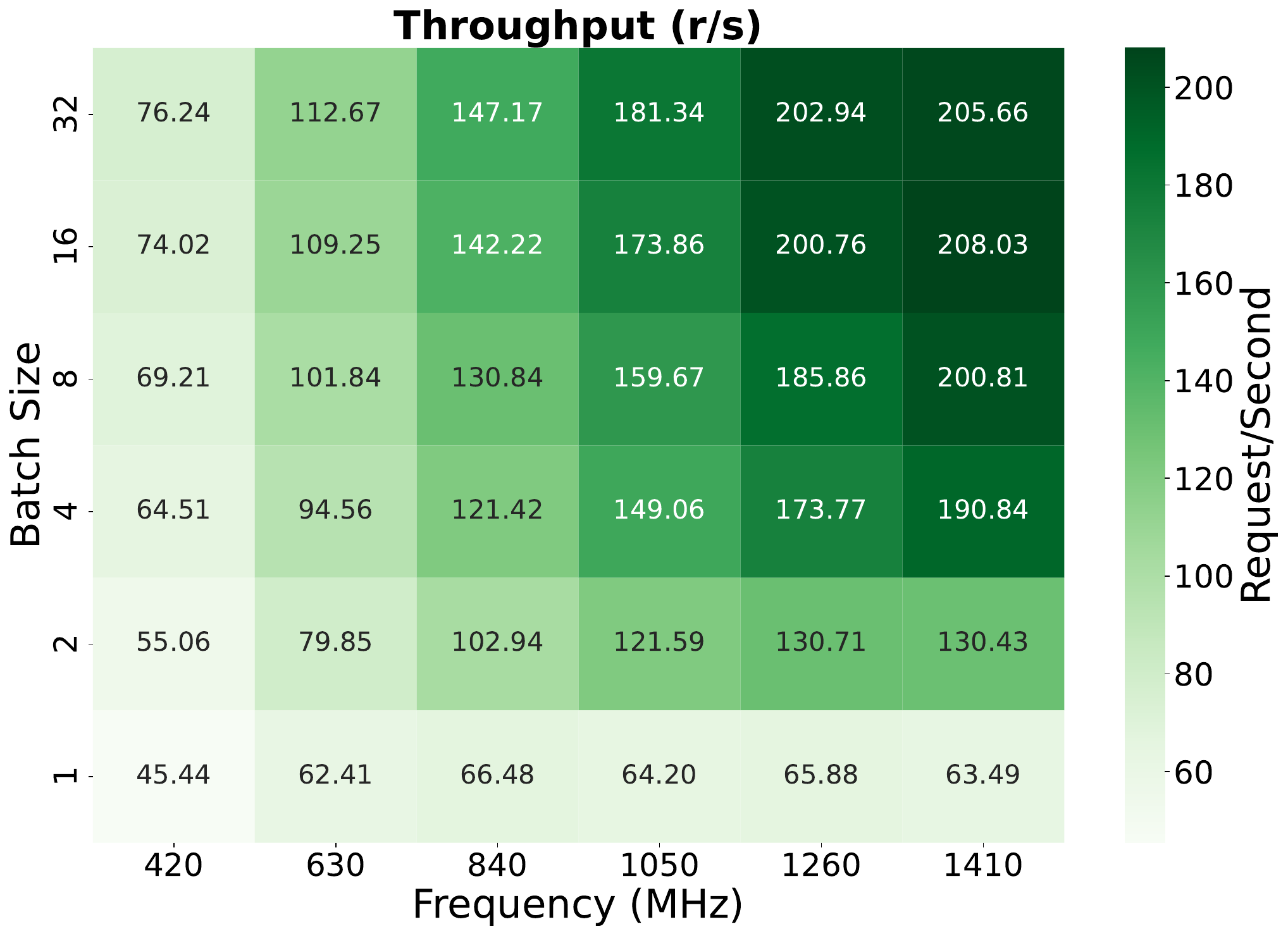}
\end{subfigure}

\vspace{-0.2em}
\centerline{\footnotesize\textbf{InternVL3 – Encode}}

\vspace{0.3em}

\begin{subfigure}{0.30\linewidth}
    \includegraphics[width=\linewidth]{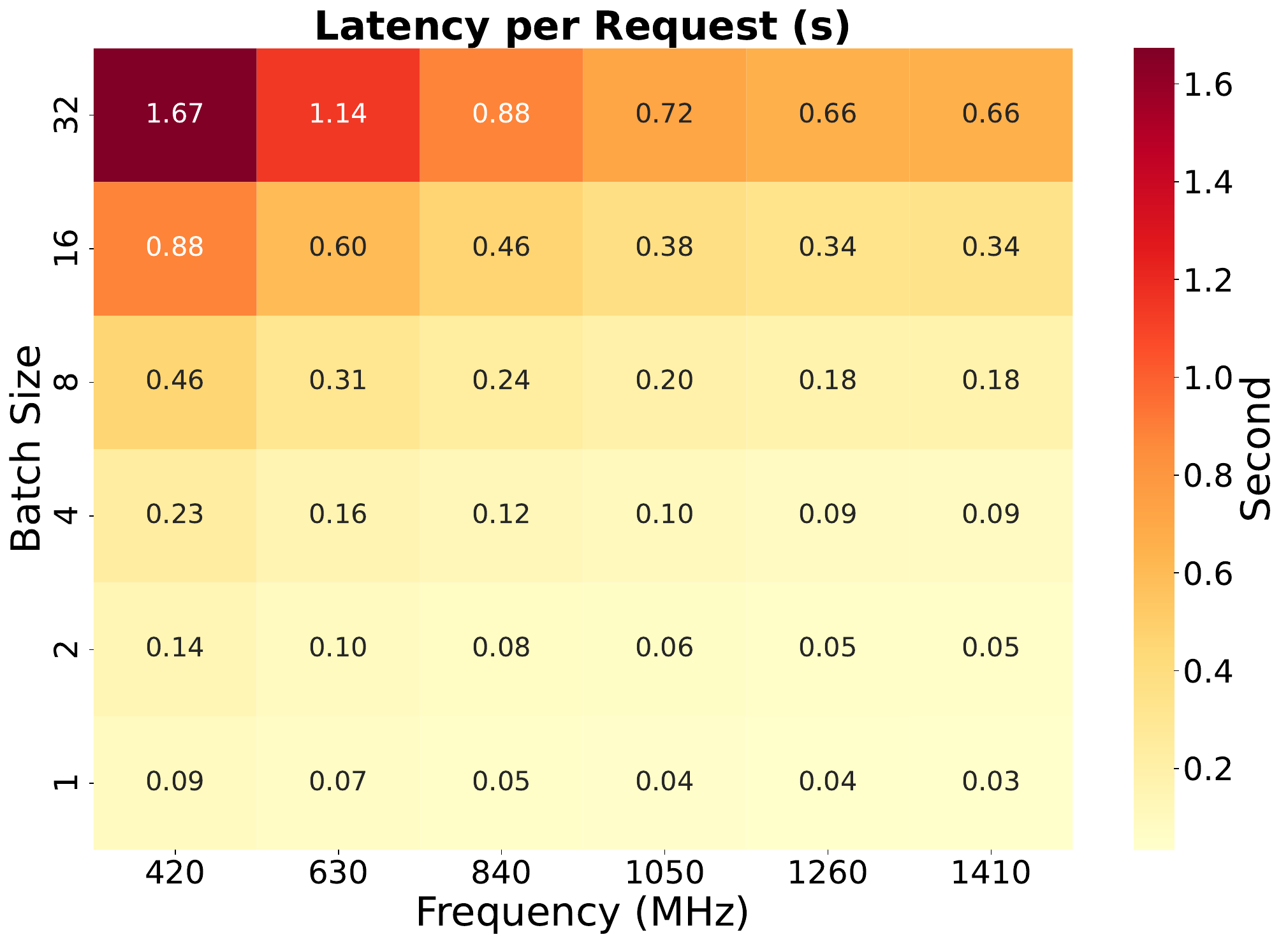}
\end{subfigure}\hfill
\begin{subfigure}{0.30\linewidth}
    \includegraphics[width=\linewidth]{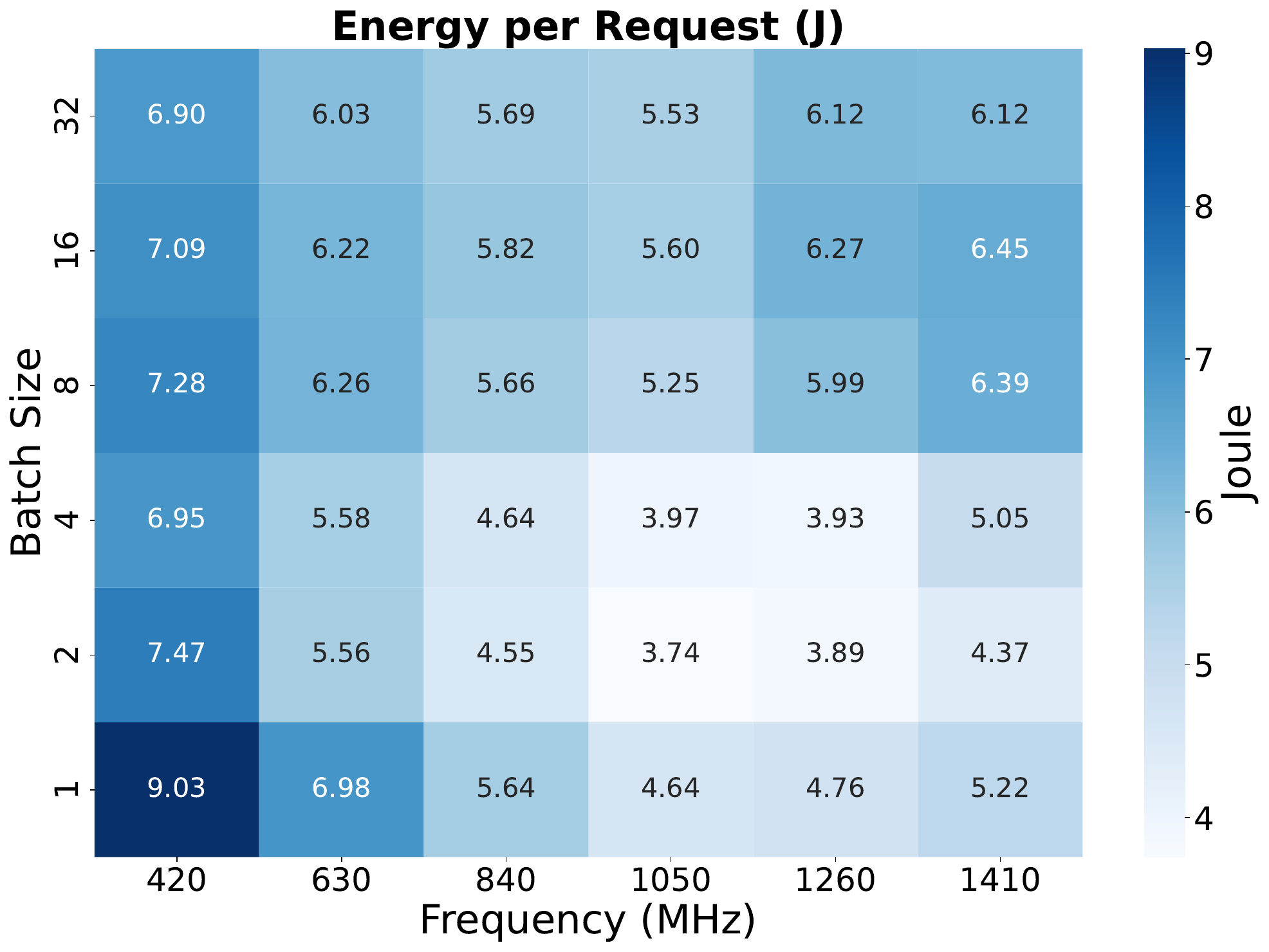}
\end{subfigure}\hfill
\begin{subfigure}{0.30\linewidth}
    \includegraphics[width=\linewidth]{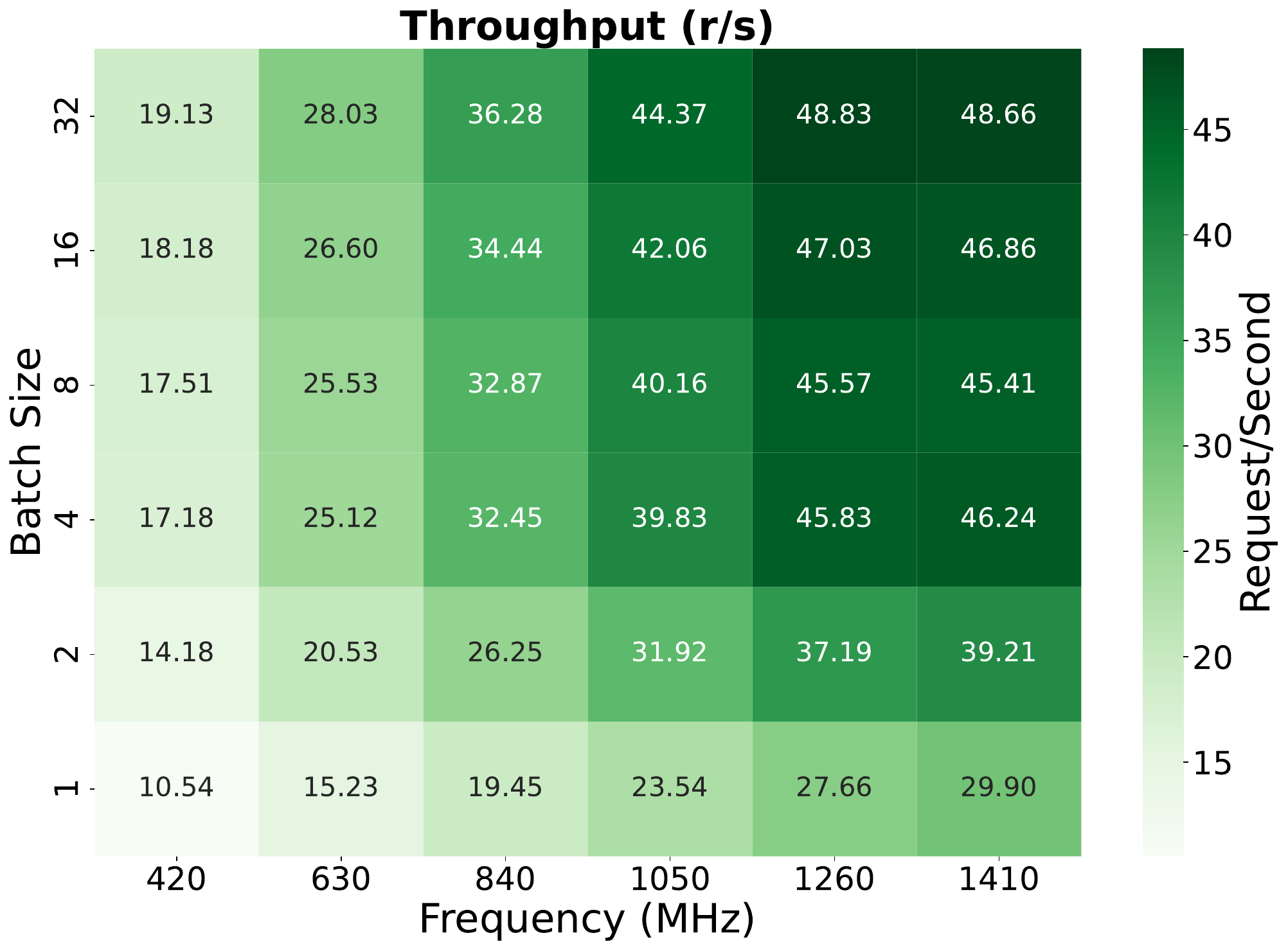}
\end{subfigure}

\vspace{-0.2em}
\centerline{\footnotesize\textbf{InternVL3 – Prefill}}

\vspace{0.6em}

\begin{subfigure}{0.30\linewidth}
    \includegraphics[width=\linewidth]{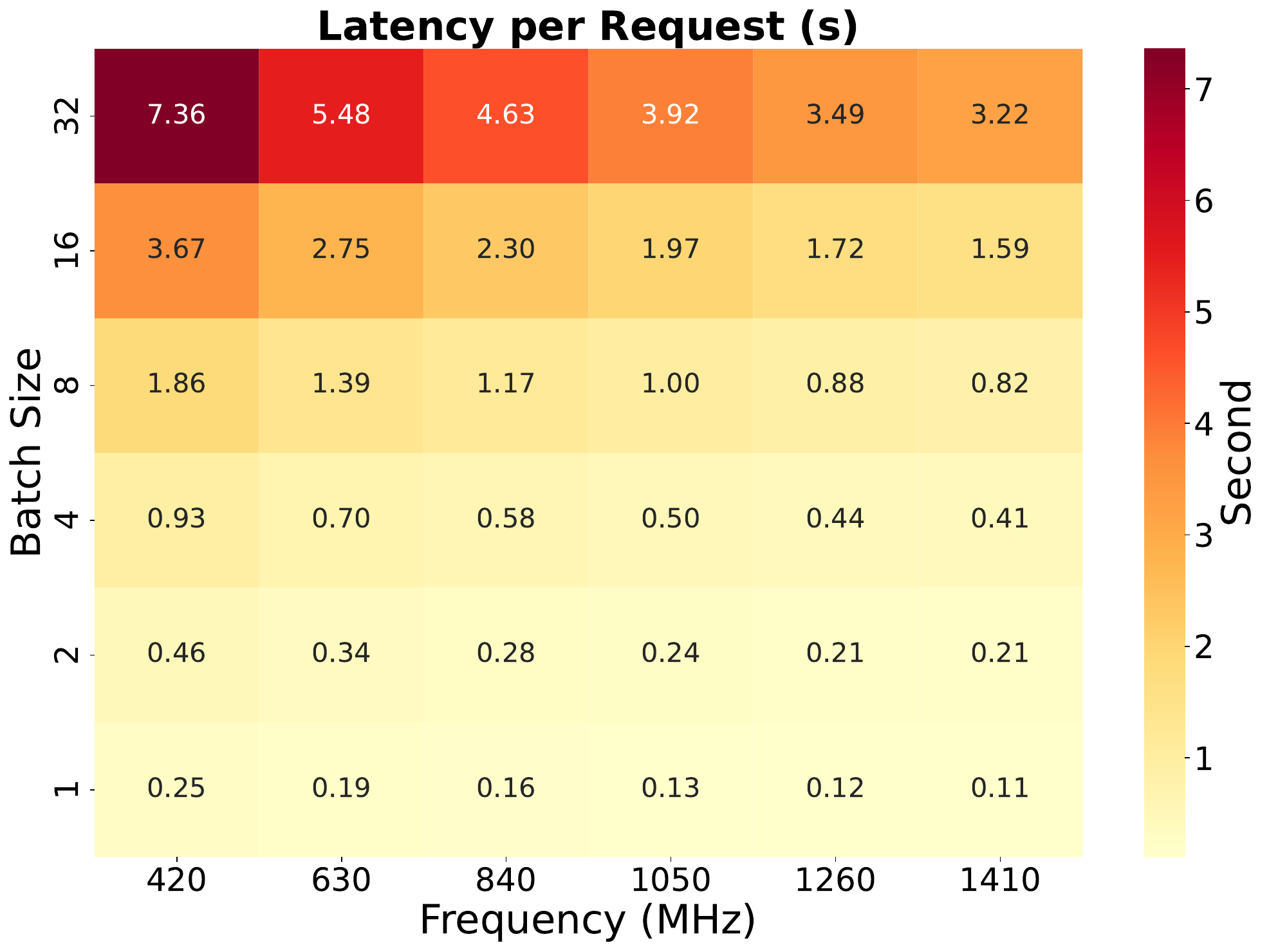}
\end{subfigure}\hfill
\begin{subfigure}{0.30\linewidth}
    \includegraphics[width=\linewidth]{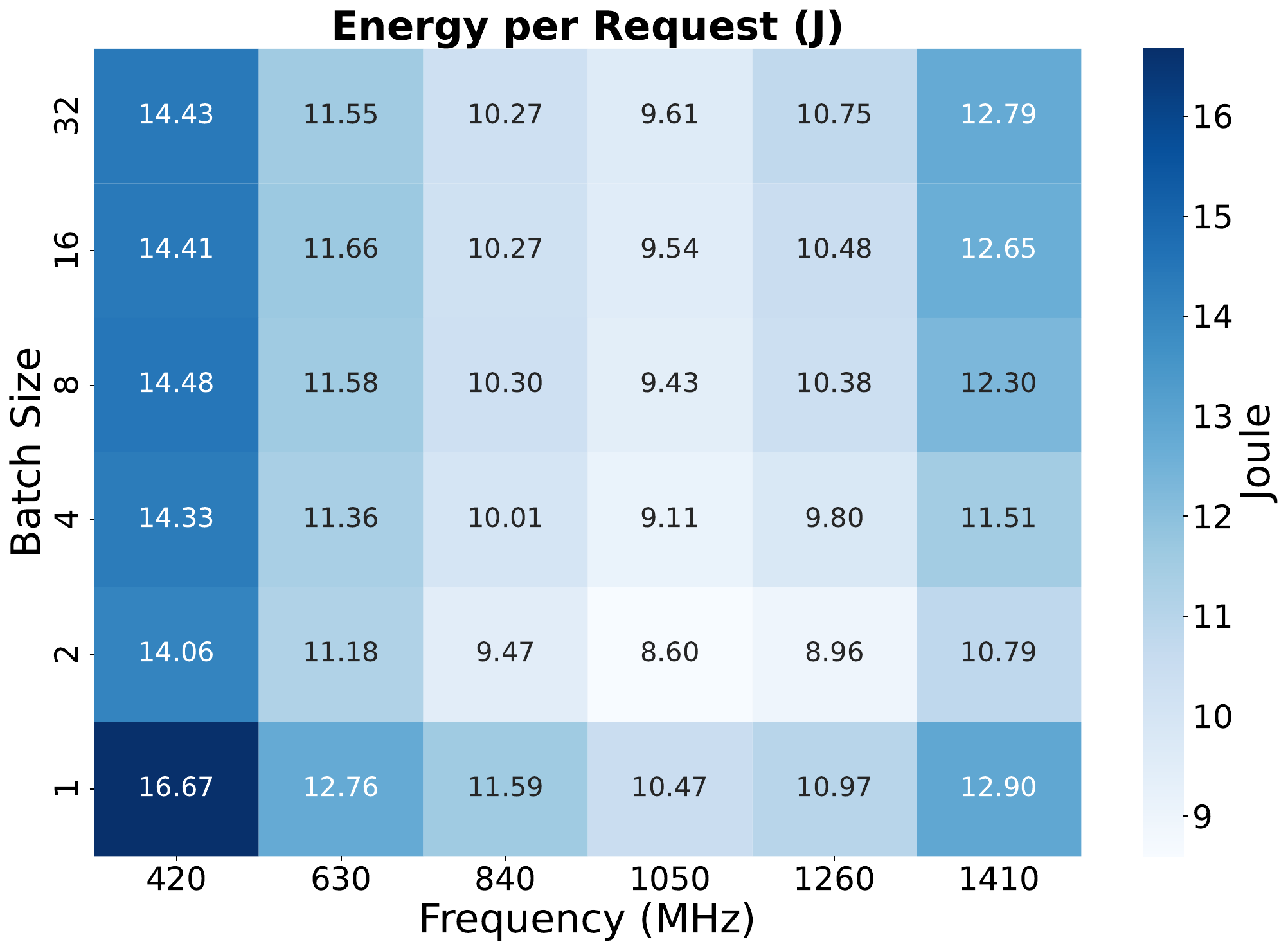}
\end{subfigure}\hfill
\begin{subfigure}{0.30\linewidth}
    \includegraphics[width=\linewidth]{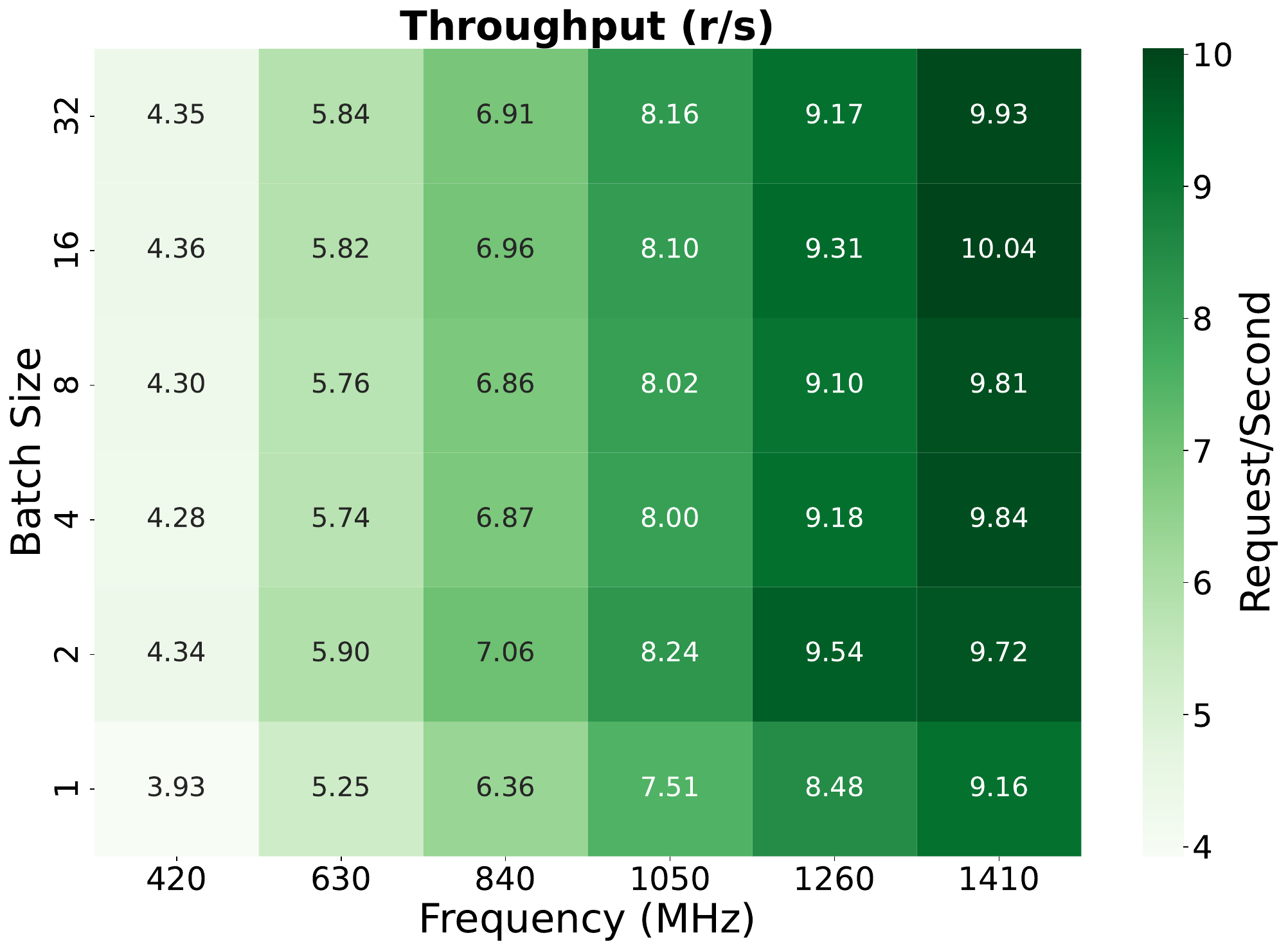}
\end{subfigure}

\vspace{-0.2em}
\centerline{\footnotesize\textbf{Qwen2.5-VL – Encode}}

\vspace{0.3em}

\begin{subfigure}{0.30\linewidth}
    \includegraphics[width=\linewidth]{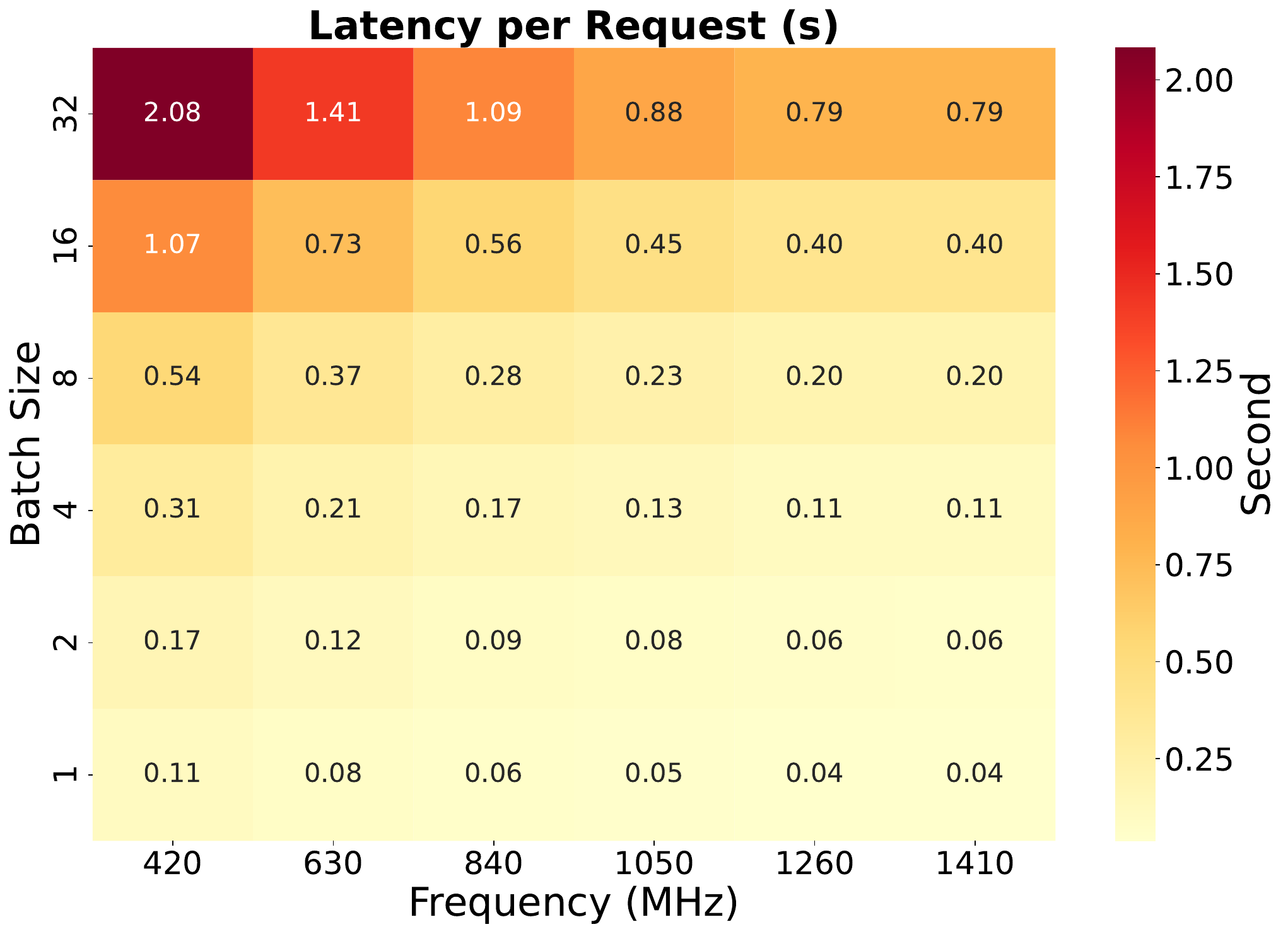}
\end{subfigure}\hfill
\begin{subfigure}{0.30\linewidth}
    \includegraphics[width=\linewidth]{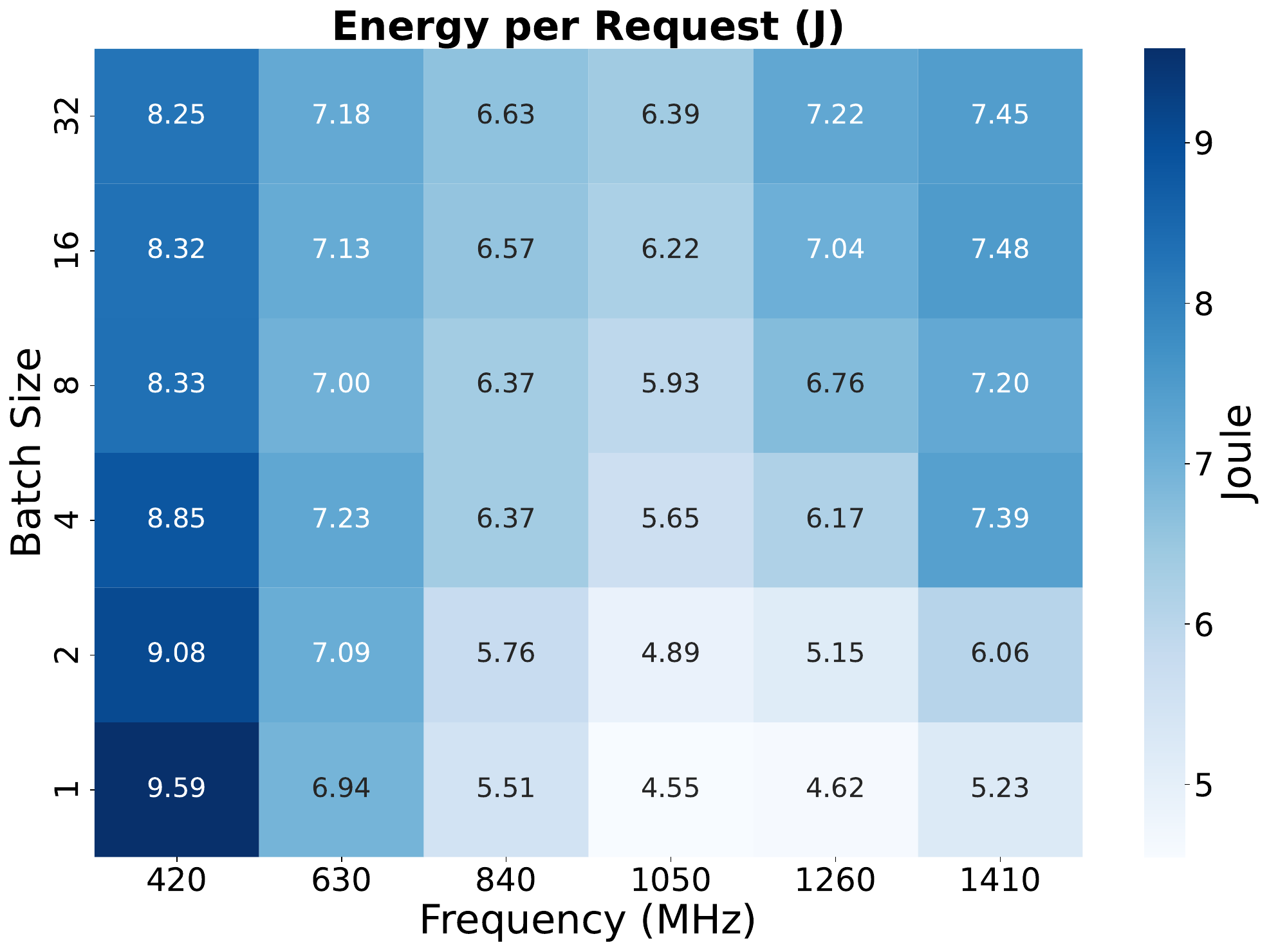}
\end{subfigure}\hfill
\begin{subfigure}{0.30\linewidth}
    \includegraphics[width=\linewidth]{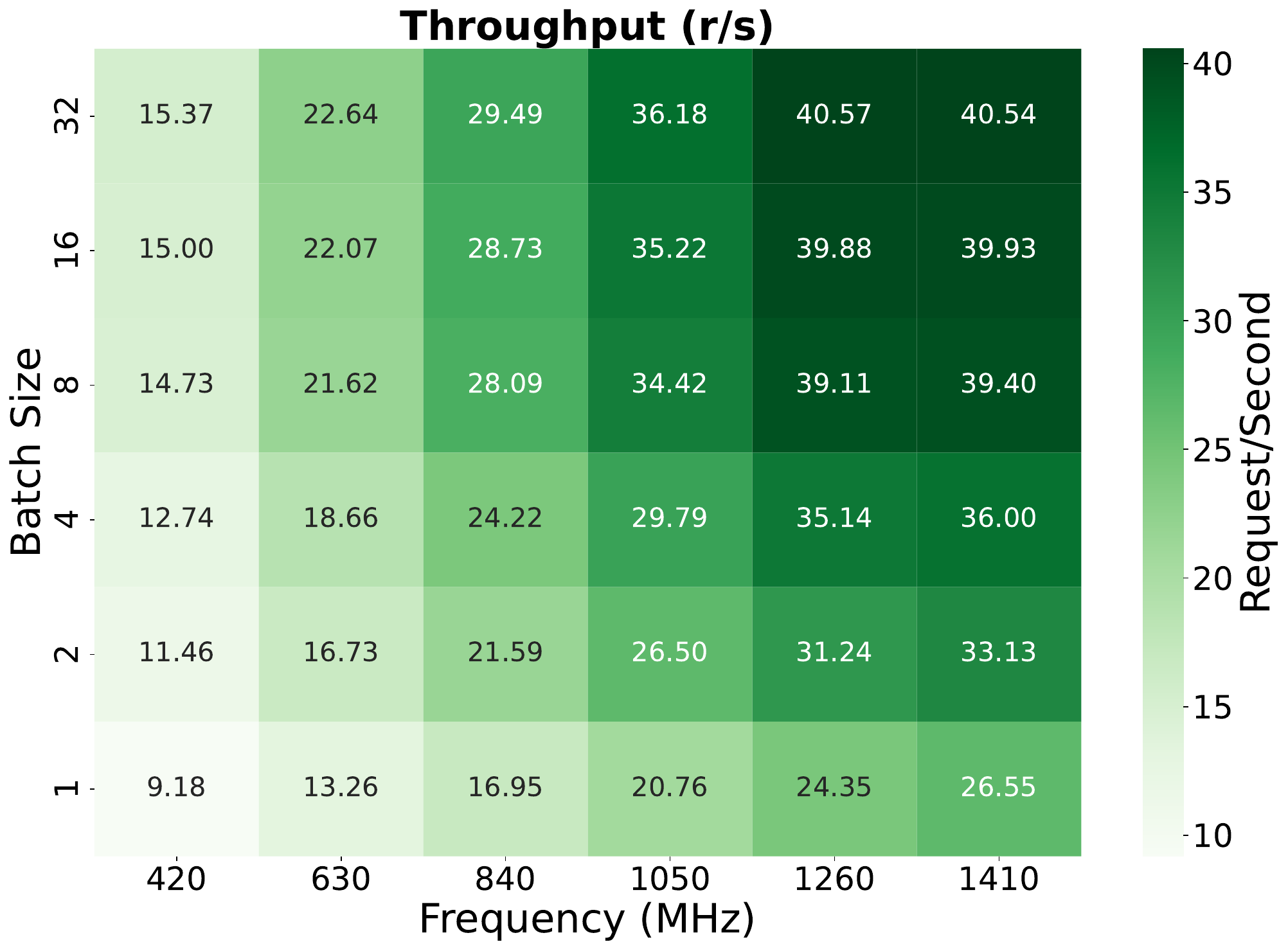}
\end{subfigure}

\vspace{-0.2em}
\centerline{\footnotesize\textbf{Qwen2.5-VL – Prefill}}

\caption{DVFS frequency–batch size heatmaps characterizing per-request energy behavior in the encoding and prefill stages of multimodal LLM inference.}
\label{fig:dvfs}
\end{figure}

%% file: sections/6-futureWork.tex
\section{Future Work}
Our study opens several promising directions for future work. 
First, while our evaluation spans diverse vision encoder and tokenization strategies, all models considered use LLM backbones in the 7B–8B parameter range. Extending this analysis to significantly smaller or larger models would help assess how model scale interacts with modality inflation and stage-level energy behavior.

Second, our current study focuses on offline characterization and static DVFS configurations. A natural next step is to design a dynamic, real-time DVFS mechanism for MLLM serving, in which GPU frequency is adaptively adjusted based on the active inference stage and anticipated input characteristics. Such a system could translate the stage-level insights from this work into practical, SLO-aware energy optimization policies.

Third, we assume a monolithic GPU-based execution pipeline. As multimodal serving systems increasingly explore disaggregated architectures, it is important to investigate how modality inflation and stage-level energy trends manifest in these settings. This includes understanding energy efficiency when different inference stages or modalities are mapped to separate accelerators.

Finally, while this work focuses on vision–language models, modern multimodal LLMs support a growing set of modalities, including audio, video, and other temporal inputs. Extending our characterization and optimization framework to these modalities remains an important open direction.

%% file: sections/9-conclusion.tex
\section{Conclusion}
In this work, we present a detailed energy characterization of vision–language LLM inference, focusing on the effects of modality inflation, inference stages, and input characteristics on GPU energy consumption. Across representative MLLMs, we find that the same multimodal inputs lead to widely divergent energy consumption, with modality-induced overheads spanning 17\% to 94\% across different model architectures. Stage-level analysis reveals that vision encoding can dominate energy consumption for certain architectures, with encoder energy exceeding 20 J and reaching over 6× that of more balanced designs, while extreme token expansion in the prefill stage can increase energy and latency by up to 12× and 8.5×, respectively. Furthermore, multi-image workloads exhibit markedly different scalability behaviors across models, with marginal energy costs varying from approximately 15 J/image to 35 J/image. These findings highlight significant opportunities for improving energy efficiency in MLLM serving through workload- and stage-wise system designs, including power management techniques such as DVFS. Overall, this work represents an initial step toward principled, architecture-aware energy optimization for multimodal large language models.